\documentclass[lettersize,journal]{IEEEtran}
\usepackage{amsmath,amsfonts}
\usepackage{algorithmic}
\usepackage{algorithm}
\usepackage{array}
\usepackage[caption=false,font=normalsize,labelfont=sf,textfont=sf]{subfig}
\usepackage{textcomp}
\usepackage{stfloats}
\usepackage{url}
\usepackage{verbatim}
\usepackage{graphicx}
\usepackage{cite}
\usepackage{svg}
\usepackage{cuted}
\usepackage{amssymb}
\usepackage{algorithm}
\usepackage{algorithmic}
\usepackage{bm} 
\usepackage{amsthm,mathtools,mathrsfs}
\usepackage{braket}
\usepackage{adjustbox}

\usepackage{makecell} 
\DeclareMathOperator*{\argmax}{arg\,max}

\usepackage{tikz}
\usetikzlibrary{positioning}
\usetikzlibrary{calc} 
\usetikzlibrary{shapes.geometric} 
\usetikzlibrary{arrows.meta}
\pgfdeclarelayer{background}
\pgfsetlayers{background,main}

\usepackage{orcidlink}

\usepackage{hhline}
\usepackage{multirow}

\theoremstyle{remark}

\begin{document}

\title{Noncoherent MIMO in the FR3 Upper Mid-Band Using Grassmannian Signaling: Prototype and Experimental Evaluation}

\author{Álvaro Pendás-Recondo\orcidlink{0000-0002-8029-8024}, Enrique Pendás-Recondo \orcidlink{0000-0002-0790-7372}, Álvaro F. Vaquero\orcidlink{0000-0001-7317-4122}, Marcos R. Pino\orcidlink{0000-0001-7468-3084}, and Jesús Alberto López-Fernández\orcidlink{0000-0001-7603-9591}
\thanks{This work was supported in part by Grant PID2023-146246OB-C32 (ANT4IT) funded by Spanish Government through the Ministerio de Ciencia e Innovación and the Agencia Estatal de Investigación MCIU/AEI/10.13039/501100011033/FEDER, European Union (EU); and in part by Government of Asturias under grant “Severo Ochoa” Program PA-22-BP21-116, and the Asturias Science, Competitiveness and Innovation Agency (SEKUENS) under projects IDE-2024-000693 and IDE/202S/001017, this last co-funded by the EU through the European Regional Development Fund (ERDF). The authors also acknowledge the co-funding and support provided by TheNextPangea, S.L.

Álvaro Pendás-Recondo, Álvaro F. Vaquero, Marcos R. Pino and Jesús Alberto López-Fernández are with the Signal Theory and Communications Group, Department of Electrical Engineering, Universidad de Oviedo, 33203, Gijón, Spain (e-mail: \{pendasalvaro,  fernandezvalvaro; mpino; jelofer\}@uniovi.es). Enrique Pendás-Recondo is with the Algebra Group, Department of Mathematics, Universidad de Oviedo, 33007, Oviedo, Spain (e-mail: pendasenrique@uniovi.es).}}



\maketitle

\begin{abstract}

This work presents a prototype implementation of a noncoherent multiple-input multiple-output (MIMO) system based on Grassmannian signalling, where neither the transmitter nor receiver requires channel state information (CSI), operating at a carrier frequency of 15 GHz in the frequency range 3 (FR3) upper mid-band envisioned for 6G. To the best of our knowledge, this is the first prototype implementation of Grassmannian signaling reported in the literature beyond sub-6 GHz bands. We identify the motivation of applying Grassmannian signaling, a form of unitary space-time modulation (USTM), in this band for wireless and mobile applications with low spectral-efficiency requirements but strict reliability and power-efficiency demands. With this idea in mind, we present a system design integrated with orthogonal frequency-division multiplexing (OFDM) that addresses the challenges associated with the higher carrier frequencies of FR3 while providing reliable operation across a broad range of scenarios, including urban and rural environments as well as vehicular communications. We then present the prototype implementation of the proposed design, based on software-defined radio (SDR) devices integrated with radio-frequency (RF) mixers, filters, and power amplifiers to ensure adequate operation in the targeted band. Over-the-air (OTA) experiments based on symbol error rate (SER) measurements under non-line-of-sight (NLoS) and line-of-sight (LoS) scenarios, and across different data rates, show good agreement with software simulations, thereby validating the prototype implementation.

\end{abstract}


\section{Introduction}
\IEEEPARstart{C}{hannel} state information (CSI) acquisition remains one of the most important challenges for the reliably and efficient performance of multiple-input multiple-output (MIMO) communications. This issue becomes more pronounced with the increasing numbers of antennas and higher operating frequencies considered in many modern and future wireless standards \cite{csi_mmwave}.  CSI at the receiver (CSIR) is typically obtained through the observation of pilot symbols periodically sent by the transmitter, whereas CSI at the transmitter (CSIT) is usually acquired through the reverse link, either via feedback conveying the CSI estimated at the receiver or, in reciprocal systems, also via pilot symbols transmitted by the receiver. Accurate CSIR acquisition in time-varying and dispersive channels depends on the periodicity of pilots in time and their distribution in frequency, as well as on the number of antennas, establishing a trade-off between estimation accuracy and the associated reduction in spectral efficiency due to pilot insertion. On the other hand, CSIT is considerably more challenging to acquire \cite{csi1}, \cite{csi2}. In feedback-based schemes, accuracy is limited by well-known factors such as finite-rate and quantized feedback, as well as feedback delays that can result in outdated information.  In reciprocity-based schemes, CSIT is also affected by channel aging, as well as imperfect calibration.

In this context, noncoherent MIMO, where neither the transmitter nor the receiver has CSI, presents an appealing alternative for paradigms such as ultra-reliable low-latency communications (URLLC), the Internet of Things (IoT), and vehicular communications, including uncrewed aerial vehicles (UAVs), especially for control and safety-critical links. Although different strategies for noncoherent MIMO have been proposed \cite{nonch_review}, including energy detection \cite{energy_based}, differential \cite{diff_survey}, and index modulation schemes \cite{index_mod}, we focus on unitary space-time modulation (USTM), which is based on the transmission of semi-unitary space-time matrices. While USTM can be adopted in schemes without CSIT but with CSIR \cite{gt2}, it can also enable communication without CSIR when the space-time constellation matrices are designed as Stiefel representatives of different points on the Grassmann manifold \cite{gt3}. This solution only relies on the assumption that the unknown channel matrix remains constant over a certain number of symbols, commonly referred to as the coherence block. Considering equal-energy signals, Grassmannian signaling is optimal in the sense of minimizing the union bound on the error probability under Rayleigh block fading \cite{gt3}. Remarkably, it can also achieve a significant fraction of the coherent-capacity gain at high SNR, with this fraction increasing as the channel coherence interval grows. However, a sufficiently long coherence block would also favour a pilot-based scheme, since the associated reduction in spectral efficiency would then be limited. This trade-off has already been evaluated in several comparative studies, including \cite{expmap2}, \cite{simo_grass}, and \cite{joint_geo}. Overall, Grassmannian signaling is particularly well suited to short-packet scenarios with limited coherence blocks, corresponding to rapidly varying channels, as well as to burst transmissions, where CSI acquisition is costly and/or unreliable.

Motivated by these aspects, many recent works have investigated the design of Grassmannian constellations for different scenarios, including single-input multiple-output (SIMO) systems \cite{cubesplit}, \cite{simo_grass}, \cite{simo_grass2}, \cite{simo_nograss1}, \cite{simo_nograss2}; MIMO systems with maximum-likelihood (ML) detection targeting minimum symbol error rate (SER) and bit error rate (BER) \cite{ub}, \cite{do2}, \cite{do1}, \cite{dp}; MIMO systems with reduced-complexity detection \cite{expmap2}, \cite{expmap1}, \cite{struct_can}; and multiuser MIMO systems \cite{joint_geo}, \cite{grassma}.

\subsection{Grassmannian and USTM Prototype Implementations}

Although Grassmannian signaling and USTM for noncoherent MIMO have been extensively studied theoretically, and many constellation designs are available, very few experimental prototypes have been reported in the literature. Table \ref{table_imp} presents previous prototype implementations, indicating their year of publication, carrier frequency, and architecture. The work in \cite{exp1} presented an initial experimental evaluation at $220$ MHz using a single-carrier architecture with a bandwidth of $4$ kHz, whereas \cite{exp4} explored the specific application of Grassmannian signaling to low-probability-of-detection communications in a single-input single-output (SISO) scenario. More closely related to our work are \cite{exp2} and \cite{exp3}, which are based on MIMO-OFDM architectures, although operating in sub-6-GHz bands. Whereas the works in \cite{exp1}, \cite{exp4}, and \cite{exp2} correspond to conference publications, \cite{exp3} is a doctoral thesis on the design of Grassmannian constellations, which led, among others, to the works in \cite{simo_grass} and \cite{ub}, and includes a chapter dedicated to the experimental validation of those constellations. To the best of our knowledge, this is the first Grassmannian or USTM prototype implementation beyond sub-6-GHz bands, as well as the first journal paper specifically dedicated to the design, implementation, and evaluation of such a prototype, with the specific contributions being described in Section \ref{cont}. Beyond validating a specific family of Grassmannian constellations, our goal is to present a complete physical-layer design capable of reliable operation across a broad range of scenarios, in accordance with the motivation described in Section \ref{mot}.

\subsection{Motivation: Noncoherent MIMO in the FR3 Upper Mid-Band}\label{mot}

Another aspect that has not been extensively explored to date is the full potential of Grassmannian schemes as the operating frequency increases. For SISO systems, the interest in using schemes such as OOK, ASK, FSK, and DPSK to avoid phase-coherence requirements, particularly in sub-THz and THz communications, is well known. Extending some of these strategies to MIMO naturally leads to several of the aforementioned approaches based on energy detection \cite{energy_2} and differential noncoherent MIMO \cite{diff_2}.

From the perspective of USTM and Grassmannian signaling, a complex trade-off involving multiple variables arises. As the operating frequency increases, for instance, from sub-6 GHz bands to mmWave or frequency range 2 (FR2) bands in 5G (24.25 - 71 GHz) or to sub-THz bands (roughly up to 300 GHz), the cost and hardware complexity associated with a larger number of antennas also increase, with one of the main limiting factors being the number of independent digital chains that can provide MIMO diversity and multiplexing gains beyond analog beamforming. Nevertheless, most works on Grassmannian signaling have demonstrated its potential for relatively small numbers of antennas, ranging from SIMO schemes \cite{cubesplit}, \cite{simo_grass} to MIMO systems with two to four transmit and receive antennas (independent digital chains) \cite{gd}. Furthermore, the Grassmannian approach inherently removes phase-coherence constraints between chains, which significantly simplifies hardware implementation, as demonstrated in the presented prototype. Finally, in FR2 bands, accurate CSI acquisition is considerably more challenging than in conventional sub-6 GHz bands \cite{csitm}. Beyond hardware limitations, this difficulty is fundamentally linked to the shorter wavelengths, which cause even small changes in receiver position to potentially induce significant variations in the channel matrix.

While these arguments could provide a strong case for Grassmannian signaling in FR2 and beyond, several additional factors must be considered. Severe propagation and penetration losses have significantly slowed the widespread adoption of FR2 bands and have largely confined their use to line-of-sight (LoS) scenarios. This limitation has, in turn, motivated intensive research into technologies capable of establishing or enhancing LoS propagation conditions, such as intelligent reflecting surfaces (IRSs), smart electromagnetic skins (SESs) \cite{ses_mag}, and reconfigurable intelligent surfaces (RISs). If FR2 bands—and, even more so, sub-THz bands—are largely restricted to such scenarios, the discussion changes substantially. Under LoS conditions, even when established through a relay, the channel matrix tends to exhibit strong spatial and temporal correlations. This structure facilitates channel tracking at the receiver using signal processing techniques while requiring only a limited number of pilot symbols from the transmitter \cite{csi_track}, thereby shifting the interest back toward conventional pilot-based schemes. The same applies to many use cases in sub-6 GHz bands, where, even under non-line-of-sight (NLoS) conditions, temporal channel variations may not be sufficiently rapid to result in very short coherence blocks (a few symbols) and, consequently, to justify the use of Grassmannian signaling.

In this context, the envisioned frequency range 3 (FR3) band for 6G, spanning approximately 7-24 GHz and also referred to as the upper mid-band, has emerged as a promising trade-off between implementation complexity and available spectrum resources, potentially enabling a broader range of use cases and deployment scenarios than FR2. Although substantial work remains to be done, measurement campaigns in FR3 have already demonstrated clear advantages in terms of propagation and penetration losses over FR2 under NLoS conditions \cite{fr3_nlos}. This opens a new potential use case for Grassmannian signaling, namely MIMO wireless communications with a relatively small number of antennas operating in FR3 under rapidly and unpredictably varying channel conditions. Examples include UAV control links, as well as vehicle-to-vehicle (V2V) and vehicle-to-everything (V2X) communications in scenarios where LoS conditions cannot be guaranteed. In this context, Grassmannian signaling offers a reliable solution that can achieve considerable power efficiency in terms of SER and BER compared to coherent schemes, although limited to operation in low-spectral-efficiency regimes (e.g., in the range of 0.25 - 1 bps/Hz for the constellations proposed in \cite{gd} and adopted in this work), mostly due to the computational complexity of ML decoding. However, for URLLC use cases, where such data rates are common, and given the abundant spectrum available in FR3, this limitation in spectral efficiency may not represent a significant drawback.

\begin{table}[t]
\caption{Grassmannian and USTM Prototype Implementations}
\centering
\begin{tabular}{|c|c|c|c|c|}
\hline
Reference & Year & Frequency & Architecture \\ \hline \hline

\cite{exp1} & 2004 & $220$ MHz & Single-Carrier MIMO  \\ \hline

\cite{exp2} & 2017 & $2.487$ GHz & MIMO-OFDM   \\ \hline

\cite{exp3} & 2024 & $1$ GHz & MIMO-OFDM  \\ \hline

\cite{exp4} & 2025 & $2.5$ GHz & Spread spectrum SISO  \\ \hline

\textbf{This work} & 2026 & $15$ GHz & MIMO-OFDM   \\ \hline

\end{tabular}
\label{table_imp}
\end{table}

\subsection{Contributions}\label{cont}
The main contributions of this work are summarized as follows:

\begin{itemize}
    \item We present and evaluate a noncoherent MIMO prototype based on the combination of Grassmannian signaling and orthogonal frequency-division multiplexing (OFDM), operating in the FR3 upper mid-band at a carrier frequency of $f_{\mathrm{c}}=15$ GHz, following the motivations described in Section \ref{mot}. To the best of our knowledge, this is the first prototype implementation of Grassmannian signaling beyond the sub-6 GHz or frequency range 1 (FR1) bands.

    \item Focusing on potential use cases for noncoherent MIMO based on Grassmannian signaling, we design the system to operate under restrictive conditions while preserving substantial flexibility across different communication scenarios, as may be required in control-link and safety-critical applications. Specifically, the design must support both LoS and NLoS conditions and a very stringent root mean square (RMS) delay spread of $\tau_{\mathrm{rms}}=955$ ns, as specified for the 3GPP urban macrocell (UMa) scenario at $15$ GHz in 3GPP TR 38.901. Regarding mobility, the design accounts for Doppler shifts up to a maximum velocity of $v_{\mathrm{max}}=35$ m/s ($126$ km/h) at both the transmitter and the receiver. Nevertheless, the rationale behind each design choice, including the subcarrier spacing, space-time matrix dimensions, and number of subcarriers, is detailed to facilitate straightforward adaptation to other scenarios.

    \item We present the prototype implementation of the proposed system, based on software-defined radio (SDR) devices and digital signal processing blocks that perform ad hoc time and frequency synchronization, as well as symbol decoding. The SDR devices are integrated with radio-frequency (RF) mixers, power amplifiers, and filters to ensure adequate operation in the FR3 band.
    
    \item We present over-the-air (OTA) experimental results for the prototype under both NLoS (Rayleigh block-fading) and LoS (strong dominant-path) scenarios. The close agreement between the measured SER across different propagation scenarios and constellation sizes (data rates) and software simulations validates the prototype implementation.

    \item Although the prototype is restricted to specific design choices, partly due to hardware limitations, such as a bandwidth of $5$ MHz and two transmit and two receive antennas, we also discuss how the proposed design can be adapted to other scenarios. For instance, increasing the bandwidth is straightforward due to the integration with OFDM, provided that the required sampling rate is supported and the design parameters are adjusted according to the resulting symbol duration. Similarly, increasing the number of antennas is straightforward if additional digital and hardware chains are available, particularly because Grassmannian signaling removes the requirement of phase coherence across RF chains imposed by most MIMO systems. However, as the dimensions of the space-time matrices increase, the computational complexity of ML decoding may become the limiting factor as the number of constellation points grows.
    
\end{itemize}

The remainder of this paper is organized as follows. Section \ref{SM} introduces the system model for noncoherent MIMO communications with Grassmannian signaling and its integration with OFDM. Section \ref{SD} describes the system design according to the considered input parameters and specifications and discusses possible adaptations to different scenarios. The prototype implementation is presented in Section \ref{PI}. The OTA experimental results and their comparison with software simulations are presented in Section \ref{EE}. Finally, conclusions are drawn in Section \ref{C}.

\textit{Notation}: Scalars are denoted by italic letters (either lowercase or uppercase, $x$, $X$), while matrices are denoted by bold uppercase letters, $\mathbf{X}$. The transpose, conjugate transpose, determinant, and trace of a matrix are denoted by $\left(\cdot\right)^T$, $\left(\cdot\right)^H$, $\operatorname{det}\left(\cdot\right)$, and $\operatorname{tr}\left(\cdot\right)$, respectively. The $ M$-dimensional identity matrix is denoted by $\mathbf{I}_M$. Subspaces and sets are denoted by calligraphic letters, $\mathcal{X}$. The imaginary unit is denoted as $j$. The absolute value of a scalar, or the cardinality of a set, is denoted by $\left|\cdot\right|$. The floor of a scalar $X$, denoted as $\lfloor X \rfloor$, is defined as the unique integer $n \in \mathbb{Z}$ such that
$n \le X < n + 1$. The expected value is denoted as $\operatorname{E}\{\cdot\}$. The Frobenius norm of a matrix is written as $\|\mathbf{X}\|_F$. A circularly-symmetric complex Gaussian distribution with mean $\mu$ and variance $\sigma^2$ is denoted by $\mathcal{CN}\left(\mu,\sigma^{2}\right)$.

\section{System Model}\label{SM}
\subsection{Grassmannian Signaling Preliminaries}\label{GSP}
We first present a very brief introduction to the mathematical background for Grassmannian signaling. A more detailed discussion in the context of noncoherent MIMO communications can be found in \cite{gd}.

The {\em complex Grassmann manifold} or {\em complex Grassmannian} $\textup{Gr}_{\mathbb{C}}(T,M)$ is defined as the set of all $M$-dimensional complex subspaces $\mathcal{U}$ of $\mathbb{C}^T$. Each such subspace $\mathcal{U}$ can be represented by a matrix $\mathbf{U} \in \mathbb{C}^{T\times M}$ whose $M$ columns form an orthonormal basis of $\mathcal{U}$, therefore satisfying $\mathbf{U}^H\mathbf{U}=\mathbf{I}_M$. We write in this case $\mathcal{U}=\operatorname{span}(\mathbf{U})$. These matrices are elements of the {\em complex Stiefel manifold}
\begin{equation*}
    \textup{St}_{\mathbb{C}}(T,M) \coloneqq \left\{\mathbf{U} \in \mathbb{C}^{T\times M}: \mathbf{U}^H\mathbf{U}=\mathbf{I}_M \right\}.
\end{equation*}
Although the matrix $\mathbf{U} \in \textup{St}_{\mathbb{C}}(T,M)$ completely determines the subspace $\mathcal{U}=\operatorname{span}(\mathbf{U})$, this representation is not unique: $\mathbf{U}, \mathbf{\tilde{U}} \in \textup{St}_{\mathbb{C}}(T,M)$ satisfy $\operatorname{span}(\mathbf{U})=\operatorname{span}(\mathbf{\tilde{U}})$ if and only if $\mathbf{\tilde{U}}=\mathbf{U}\mathbf{R}$, where $\mathbf{R}$ is an element of the {\em unitary group}
\begin{equation*}
    U(M) \coloneqq \left\{\mathbf{R} \in \mathbb{C}^{M\times M}: \mathbf{R}^H\mathbf{R}=\mathbf{I}_M=\mathbf{R}\mathbf{R}^H \right\}.
\end{equation*}
All matrices that generate the same subspace $\mathcal{U}=\operatorname{span}(\mathbf{U})$ are denoted as $[\mathbf{U}]$ and are called the {\em equivalence class} of $\mathbf{U}$:
\begin{equation*}
    [\mathbf{U}] \coloneqq \left\{\mathbf{U}\mathbf{R} \in \textup{St}_{\mathbb{C}}(T,M): \mathbf{R} \in U(M)\right\}.
\end{equation*}
In this way, each subspace $\mathcal{U}=\operatorname{span}(\mathbf{U})$ uniquely corresponds to $[\mathbf{U}]$, so we can make the identification $[\mathbf{U}]\equiv \mathcal{U}$ and say that $[\mathbf{U}]$ is a {\em point} in $\textup{Gr}_{\mathbb{C}}(T,M)$, whereas the matrix $\mathbf{U} \in \textup{St}_{\mathbb{C}}(T,M)$ is called a {\em Stiefel representative} of $\mathcal{U}$.

Given any $\mathbf{\tilde{U}} \in [\mathbf{U}]$, the matrix $\mathbf{P}_{[\mathbf{U}]} = \mathbf{\tilde{U}}\mathbf{\tilde{U}}^H \in \mathbb{C}^{T\times T}$ is called the {\em orthogonal projector} onto the subspace $\mathcal{U}=\operatorname{span}(\mathbf{U})$, i.e., it is the unique matrix satisfying $\mathbf{P}_{[\mathbf{U}]}^H=\mathbf{P}_{[\mathbf{U}]}$, $\mathbf{P}_{[\mathbf{U}]}^2=\mathbf{P}_{[\mathbf{U}]}$, and $\operatorname{range}(\mathbf{P}_{\mathbf{[U]}})=\mathcal{U}$. This projector is independent of the chosen Stiefel representative, so it also provides a unique characterization of each point $[\mathbf{U}] \in \textup{Gr}_{\mathbb{C}}(T,M)$.

The key intuition behind Grassmannian signaling for noncoherent communications is that, given an unknown channel matrix $\mathbf{H}$, each column of $\mathbf{U}\mathbf{H}$ is a linear combination of the columns of $\mathbf{U}$, such that $\operatorname{span}(\mathbf{U}\mathbf{H})\subseteq\operatorname{span}(\mathbf{U})$. Hence, although the unknown channel alters the received matrix, the received signal in the absence of noise remains in the transmitted subspace associated with the Grassmannian point $[\mathbf{U}]$.

\subsection{Narrowband Channel Model}\label{NCM}
We first introduce Grassmannian signaling considering a narrowband channel model, as this is the common assumption for ease of exposition \cite{nonch_review}. Consider a MIMO wireless communication system with $M$ transmit antennas and $N$ receive antennas over a frequency-flat block-fading channel with a coherence time of $T_{\mathrm{ch}}$ symbols, where $T_{\mathrm{ch}} \geq 2M$. That is, the channel matrix $\mathbf{H} \in \mathbb{C}^{M \times N}$ is assumed to remain constant over $T_{\mathrm{ch}}$ symbols, although its entries are unknown to both the transmitter and the receiver. Within a block of $T\leq T_{\mathrm{ch}}$ symbols, the transmitter sends a unitary space-time matrix $\mathbf{X} \in \mathbb{C}^{T \times M}$ satisfying $\mathbf{X}^{H}\mathbf{X} = \mathbf{I}_M$. This is equivalent to saying that the columns of $\mathbf{X}$ are orthonormal, or that $\mathbf{X}$ is a Stiefel representative of a point $[\mathbf{X}]$ in the complex Grassmann manifold $\textup{Gr}_{\mathbb{C}}(T,M)$ (Section \ref{GSP}). The codebook, or constellation, of possible transmitted space-time matrices is denoted by the alphabet $\mathcal{X}$, whose cardinality is $\left|\mathcal{X}\right| = L$, where $L$ is the number of constellation points or symbols. For a given constellation, the communication rate is given by $R=\log_2(L)/T$ (bps/Hz), and each codeword (or block) carries $\log_2(L)$ bits of information. The received signal over a time block is denoted by $\mathbf{Y} \in \mathbb{C}^{T \times N}$ and is given by

\begin{equation}\label{Y}
\mathbf{Y} =\sqrt{\frac{T}{M}} \mathbf{X}\mathbf{H} + \sqrt{1/\rho}\mathbf{W},
\end{equation}  where $\mathbf{W}\in \mathbb{C}^{T \times N}$ denotes additive white Gaussian noise (AWGN), whose entries are i.i.d. with $w_{ij}\sim \mathcal{C}\mathcal{N}(0,1)$, and $\rho$ denotes the signal-to-noise ratio (SNR). 

Detection is performed at the receiver without knowledge of $\mathbf{H}$ using the following decoder \cite{gt2}

\begin{equation}\label{ml}
\hat{\mathbf{X}} = \argmax_{\mathbf{X}\in\mathcal{X}} \operatorname{tr}\left( \mathbf{Y}^H \mathbf{X}\mathbf{X}^H\bold{Y}\right) = \argmax_{\mathbf{X}\in\mathcal{X}} \|\mathbf{X}^H \mathbf{Y} \|_F^2,
\end{equation} where $\mathbf{X}\mathbf{X}^H \coloneqq \mathbf{P}_{[\mathbf{X}]}$ is the projector onto the subspace $\operatorname{span}(\mathbf{X})$. Importantly, this corresponds to ML detection when $\mathbf{H}$ is isotropically distributed, as in Rayleigh fading distribution, i.e., when the MIMO channel does not favor any particular spatial directions. However, even in the more realistic case where $\mathbf{H}$ favors certain spatial directions, exploiting this feature at either transmission or detection would require at least statistical CSIT or CSIR, respectively. Although statistical CSI is more feasible to obtain than perfect knowledge of the entries of $\mathbf{H}$, the favored spatial directions, i.e., its singular vectors, are also expected to vary rapidly in some of the scenarios considered in this work. Furthermore, as shown in Section \ref{NLoS}, even though it may not be optimal for all possible channel distributions, the proposed decoder performs well across all scenarios, including LoS (rank-deficient) and NLoS conditions.

\subsection{Broadband Channel Model}\label{BCH}
For a broadband channel model, we consider the integration of the narrowband system with OFDM using $K$ orthogonal frequency subcarriers. The goal is to ensure an approximately flat channel response within each subcarrier and to mitigate inter-symbol interference (ISI) caused by multipath propagation, as well as inter-carrier interference (ICI), through cyclic prefix (CP) insertion. With this in mind, the received space-time matrix at the $k$-th subcarrier, $k\in[0,K-1]$, can be expressed as

\begin{equation}\label{Yofdm}
\mathbf{Y}[k] =\sqrt{\frac{T}{M}} \mathbf{X}[k]\mathbf{H}[k] + \sqrt{1/\rho}\mathbf{W}[k],
\end{equation} where $K$ noncoherent communications are carried out in parallel, one per subcarrier. That is, $T$ complete OFDM symbols are required before independently decoding $K$ constellation symbols. The important implication is that, to satisfy the coherence condition, the channel must remain quasi-constant over $T$ complete OFDM symbols, including the CP duration. Finally, note that a single (average) SNR value is considered, while channel power variations across subcarriers can be represented through $\mathbf{H}[k]$.

\subsection{Discrete and Continuous Time-Frequency Parameters}\label{DCTM}
Up to this point, we have implicitly considered a discrete-time model in which both $T$ and $T_{\mathrm{ch}}$ are integers measured in symbol intervals or channel uses. Focusing on OFDM, we denote the communication bandwidth by $\mathrm{BW}$ (Hz) and the sampling period by $t_s$ (s), with $t_s = 1/\mathrm{BW}$. The subcarrier spacing is denoted by $\Delta f = \mathrm{BW}/K$ (Hz), and $K_{\mathrm{cp}}$ denotes the CP length in samples. The total OFDM symbol duration is given by $t_{\mathrm{OFDM}} = (K+K_{\mathrm{cp}})t_s$ (s), and the CP duration by $t_{\mathrm{cp}} = K_{\mathrm{cp}}t_s$ (s). The channel coherence time may be expressed either in time units as $t_{\mathrm{ch}}$ (s), or in discrete-time units relative to the OFDM symbol duration as $T_\mathrm{ch} = \lfloor t_{\mathrm{ch}}/t_{\mathrm{OFDM}} \rfloor$. The time block length $T$ represents the number of OFDM symbols over which each subcarrier transmits a codeword and can be adjusted subject to the condition $T\leq T_{\mathrm{ch}}$. The total block duration in seconds is then given by $T_{\mathrm{t}} = Tt_{\mathrm{OFDM}}$ (s).

\section{System Design}\label{SD}
In this section, we present the system design, including explicit choices of system parameters and a numerical analysis of potential channel effects. Our goal is to illustrate the flexibility of combining OFDM and noncoherent MIMO through a particular case study. However, we provide the rationale behind each decision so that the design can be readily adapted to different conditions depending on the final application. We begin by defining the input variables, where we include the number of antennas and frequency band, and then move to study the adjustment of  the remaining system variables to achieve the desired behavior.

\subsection{Input Parameters}\label{IP}

We consider $M=N=2$, i.e., both the transmitter and receiver are equipped with two antennas. We limit this prototype to a relatively small number of antennas, since two transmit and two receive antennas are sufficient to validate noncoherent MIMO operation in this frequency band. In particular, if additional antennas are incorporated at either the transmitter or the receiver, the hardware implementation requires no phase coherence, but only common frequency and clock references (Section \ref{PI}). Meanwhile, the coding and decoding strategies remain essentially unchanged, except for the dimensions of the used space-time matrices. Provided that the required hardware is available, the main consideration when increasing the number of antennas (e.g. $M=N=4$) is that the computational complexity of the ML decoder in (\ref{ml}) may become prohibitive, depending on the number of constellation points. Note that, for arbitrary Grassmannian constellations, this computational complexity is given by $\mathcal{O}(TMNL)$ \cite[Section IV-D]{gd}. In such cases, reduced-complexity decoders can be employed, such as those proposed in \cite{expmap2}, \cite{cubesplit}, and \cite{simo_grass}.

The selected carrier frequency is $f_{\mathrm{c}}=15$ GHz. This choice is motivated by the fact that it is perhaps the highest frequency at which this particular family of noncoherent communication systems may find practical applicability in terrestrial scenarios. Systems operating at higher frequencies, such as $f_{\mathrm{c}}=25$ GHz in the mmWave bands or $f_{\mathrm{c}}=100$ GHz in the sub-terahertz bands, tend to rely on LoS conditions, even when these are established through relaying technologies, to overcome the significant penetration and propagation losses associated with these frequency bands. In such cases, even under high-mobility conditions, the channel variations tend to exhibit considerable temporal correlation, simplifying CSI acquisition and reducing its associated cost. At the receiver, CSIR can be estimated and subsequently tracked through signal processing, preventing an excessive increase in pilot overhead. At the transmitter, CSIT is also less likely to become outdated. These advantages, however, arise from the initial restriction of operation to LoS conditions. Satellite communications may nonetheless require a separate analysis. Although the preceding general observations regarding LoS conditions broadly apply, the higher latency relative to terrestrial communications may severely limit the acquisition of updated CSIT, even when temporal channel correlation is present. Although beyond the scope of this work, whether noncoherent MIMO based on USTM, Grassmannian signaling or differential schemes could be beneficial for satellite communications is an interesting question for future investigation.

We assume $\mathrm{BW}=5$ MHz, so OTA prototyping and testing based on SDR are not limited by the signal processing sampling rate. However, using OFDM, the system can be readily adapted to other bandwidths. Considering these input parameters, our goal is to present a system design that supports reliable communications across a broad range of channel conditions, with a particular focus on V2X and V2V scenarios, including UAV-to-UAV and UAV-to-vehicle links. Focusing on terrestrial communications, the system should support rural and urban areas with different time-delay profiles; NLoS (dispersive-channel) and LoS (dominant-path) conditions; and both static and mobile scenarios. In particular, at $15$ GHz, we consider the most stringent RMS delay spread, $\tau_{\mathrm{rms}}$, for the 3GPP UMa scenario, where the 3GPP TR 38.901 specifies the value $\tau_{\mathrm{rms}} = 955$ ns as the NLoS 90th-percentile RMS delay spread. For mobility, we consider a maximum velocity, for both transmitter and receiver, of $v_{\mathrm{max}}=35$ m/s ($126$ km/h). Note that urban velocities are typically lower, whereas rural delay spreads are generally much shorter. Thus, our intention is to design a system capable of operating in both environments, while a final deployment can always be tailored more specifically to a particular use case. This also applies to the carrier frequency and number of antennas, which are treated here as input parameters given the research-oriented nature of the prototype.

\subsection{Explicit Choices for System Parameters}\label{EC}

\begin{table}[t]
\caption{System design parameters.}
\centering
\begin{tabular}{|c|c|c|}
\hline
\textbf{Parameter} & \textbf{Notation} &  \textbf{Value} \\ \hline \hline

Number of transmit antennas & $M$ & $2$  \\ \hline

Number of receive antennas & $N$ & $2$ \\ \hline

Time-block length & $T$ & $4$ \\ \hline

Bandwidth & $\mathrm{BW}$ & $5$ MHz \\ \hline

Carrier frequency  & $f_{\mathrm{c}}$ & $15$ GHz\\ \hline

Number of subcarriers  & $K$ & $64$ \\ \hline

CP length in samples  & $K_{\mathrm{cp}}$ & $20$  \\ \hline

\end{tabular}
\label{table1}
\end{table}

Given the conditions presented in Section \ref{IP}, we first select the values of $K$ and $K_{\mathrm{cp}}$ to ensure robustness against multipath and Doppler effects. A summary of the system design parameters is presented in Table \ref{table1}, including $K=64$ and $K_{\mathrm{cp}}=20$. This results in $\Delta f= 78.125$ kHz and $t_{\mathrm{cp}}=4$ $\mu$s. Note that, even in the worst case with $\tau_{\mathrm{rms}} = 955$ ns, we keep $t_{\mathrm{cp}}>4\tau_{\mathrm{rms}}$, a common rule of thumb for protection against ISI caused by multipath. Regarding the Doppler effect, the maximum possible Doppler shift occurs when both the transmitter and receiver move with maximum radial velocities in opposite directions, resulting in the largest relative radial velocity, yielding $f_{\mathrm{dmax}} = 2 \displaystyle \frac{v_\mathrm{max}}{3\times10^8}f_{\mathrm{c}} = 3.5$ $\mathrm{kHz}$. A conservative rule of thumb for the minimum OFDM subcarrier spacing is therefore $\Delta f \geq 20 f_{\mathrm{dmax}} = 70$ kHz which is satisfied by our choice of $\Delta f = 78.125$ kHz. Although this criterion is used as a reference to limit ICI caused by Doppler, in practice, the maximum Doppler shift would typically occur with a dominant LoS component and thus can be compensated by a frequency synchronization algorithm, as discussed in Section \ref{PI}. However, in the presence of multipath propagation, different trajectories may experience different Doppler shifts due to the different projections of the relative velocity vector onto each propagation path. Consequently, a residual Doppler spread may remain after applying a single frequency correction. Finally, we note that, although the Doppler effect is fundamentally a time/frequency scaling effect rather than a simple frequency displacement, modeling it as a frequency offset is a valid approximation in this case, since $\mathrm{BW}\ll f_{\mathrm{c}}$.

A careful analysis is required when choosing $T$, which is set to its minimum possible value for Grassmannian signaling, i.e., $T=2M=4$. First, the specific choice of $T=2M$ when $M>1$ and $M=N$ is arguably the most relevant scenario for the application of Grassmannian signaling in terms of capacity and degrees of freedom \cite[Section II-B]{gd}. Second, from a practical perspective, this choice also limits the decoding computational complexity compared with larger values of $T$. Finally, and perhaps more importantly for this work, $T$ should be kept small enough to ensure the quasi-static channel condition $T\leq t_{\mathrm{ch}}$. With the chosen parameters, we have $t_{\mathrm{OFDM}} = 16.8$ $\mu$s and $T=4 \times 16.8=67.2$ $\mu$s, during which the complex channel matrices for each subcarrier, $\mathbf{H}[k]$, should remain approximately constant. Considering the maximum radial velocity again, the relative position between transmitter and receiver can vary by up to $4.7$ mm during $67.2$ $\mu$s, so it is reasonable to assume that the channel will not experience any significant fading variations within this interval. However, the Doppler shift can cause up to $2\pi f_{\mathrm{dmax}} t_{\mathrm{OFDM}}=0.3695$ rad, or a $21.168$ degrees ($21.168^\circ$) shift in the phase of channel entries between each consecutive OFDM symbol. Again, Doppler shifts close to this maximum value would occur in the presence of maximum radial velocity and generally under LoS dominant-path conditions, which are precisely the scenarios where the Doppler effect can be significantly mitigated through single-value frequency synchronization, as implemented in Section \ref{PI}. Even when multiple Doppler shifts are present in LoS conditions, significant components would typically be clustered around the dominant path, thus producing a reduced residual effect after single-shift correction. On the other hand, in NLoS conditions, Doppler shifts may exhibit larger variations, but their magnitudes are expected to be lower even at maximum velocities, since the propagation directions are generally not aligned with the direction of maximum radial velocity. 

\begin{figure}
\centering
\includegraphics[width=0.8\linewidth]{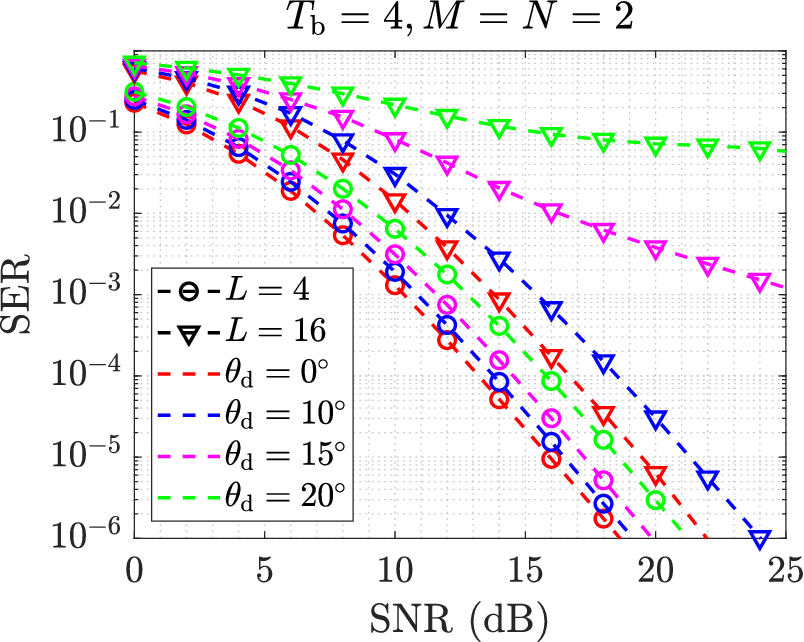}
\caption{$\mathrm{SER}$ results with $T=4$ and $M=N=2$ for $L=4$ and $L=16$, considering different values of phase shift, $\theta_{\mathrm{d}}$.}
\label{ir1}
\end{figure}

However, it is clear that this residual Doppler effect will have a non-negligible impact on the phase variations of $\mathbf{H}[k]$ across different OFDM symbols and should therefore be studied, even if only in a preliminary manner. For this purpose, we consider a constant Doppler shift $f_{\mathrm{d}}$ over the duration of the time block. Note that velocity changes, which would modify $f_{\mathrm{d}}$ over time, are negligible for $T=67.2$ $\mu$s even under extreme acceleration values. Across each block and for each subcarrier, the entries of $\mathbf{H}[k]$ experience a constant phase shift of $\theta_{\mathrm{d}} = 2\pi t_{\mathrm{OFDM}}f_{\mathrm{d}}$. Note that this phase shift is assumed to be identical across all antenna links, which can be regarded as a worst-case assumption when $\theta_{\mathrm{d}}$ is taken as its maximum value. Under this assumption, (\ref{Yofdm}) can be rewritten as

\begin{equation}\label{Yofdmth}
\mathbf{Y}[k] =\sqrt{\frac{T}{M}} \mathbf{F}\mathbf{X}[k]\mathbf{H}[k] + \sqrt{1/\rho}\mathbf{W}[k],
\end{equation} where $ \mathbf{F} = \operatorname{diag}\left(1, e^{j\theta_{\mathrm{d}}},e^{j2\theta_{\mathrm{d}}},\ldots,e^{j(T-1)\theta_{\mathrm{d}}}\right)$. Fig. \ref{ir1} displays the impact of different $\theta_{\mathrm{d}}$ values on SER performance for the chosen design $T=4,M=N=2$ with $L=4$ and $L=16$ constellation points. In this case, Rayleigh fading is assumed, so entries of $\mathbf{H}[k]$, denoted by $h_{ij}$ for the $i$-th row and $j$-th column, are independent and identically distributed (i.i.d.) and follow a complex normal distribution with zero mean and unit variance, i.e.,  $h_{ij} \sim  \mathcal{C}\mathcal{N}(0,1)$. SER results are obtained using Monte Carlo simulations using the constellations designed in \cite{gd}. It can be observed that, for the considered cases, constellation decoding exhibits remarkable robustness against phase/frequency shifts. In particular, for $L=4$, even with $\theta_{\mathrm{d}}=20^\circ$, the SER performance degrades by only about $2.5$ dB. For $L=16$, this robustness is naturally lower, although values up to $\theta_{\mathrm{d}}=10^\circ$ are still well tolerated by the system, which in our scenario corresponds to an uncorrected frequency shift of $f_{\mathrm{d}}=1.6531$ kHz. Finally, it should be noted that this phase/frequency shift may also be caused by a carrier frequency offset (CFO) arising from imperfect local oscillators, especially in low-cost hardware devices. The specific design of Grassmannian constellations for scenarios in which this effect is expected to be significant has been explored in \cite{cfo}.

\subsection{NLoS vs LoS Conditions}\label{NLoS}

Noncoherent communications are generally studied under the assumption of Rayleigh fading \cite{nonch_review}, which implicitly considers NLoS and dispersive-channel conditions. While this is a reasonable assumption given the nature of noncoherent designs, namely to operate in scenarios where CSI varies rapidly, it is also of interest to consider noncoherent schemes under LoS, especially as the frequency increases. Given fast-moving devices and short wavelengths, CSI can vary rapidly even in dominant-path scenarios, particularly due to phase variations. If trajectories do not experience abrupt variations, these phase changes can be tracked under non-dispersive conditions due to strong temporal correlation, as discussed in Section \ref{IP}. However, for trajectories with abrupt variations, as may occur, for instance, in a UAV link even under LoS conditions, this tracking may also require significant pilot overhead.

\begin{figure}
\centering
\includegraphics[width=0.8\linewidth]{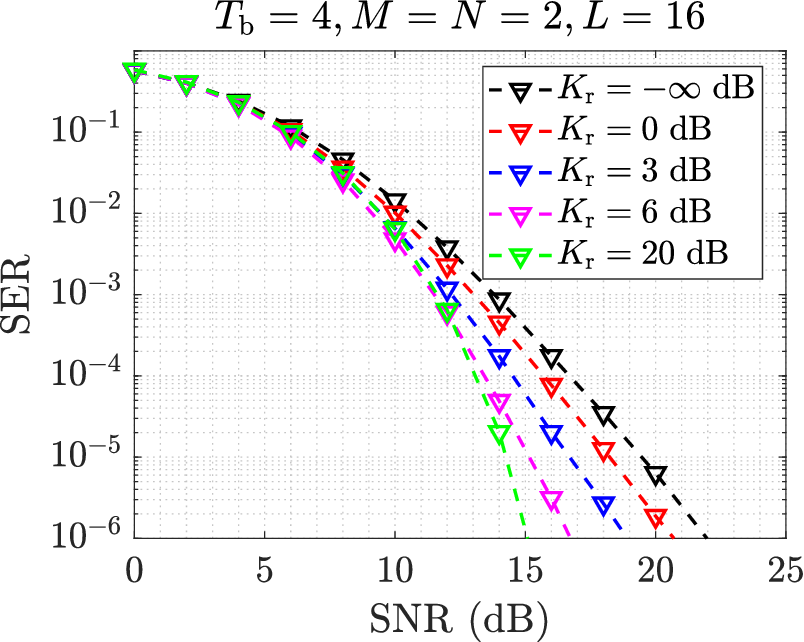}
\caption{$\mathrm{SER}$ results with $T=4$, $M=N=2$, and $L=16$ for different values of $K_{\mathrm{r}}$ (expressed in dB), which denotes the Rician $K$-factor.}
\label{fr2}
\end{figure}

Since the proposed design must be capable of operating under all the aforementioned conditions, we now numerically evaluate SER performance of the system under Rician fading, while further experimental results are reported in Section \ref{EE}. Considering Rician fading, we have

\begin{equation}\label{HRice}
\mathbf{H}[k] =\sqrt{\frac{K_{\mathrm{r}}}{K_{\mathrm{r}}+1}}\mathbf{H}_{\mathrm{LoS}}[k] + \sqrt{\frac{1}{K_{\mathrm{r}}+1}}\mathbf{H}_{\mathrm{NLoS}}[k],
\end{equation} where $K_{\mathrm{r}}$ is the Rician $K$-factor, i.e., the power ratio between the LoS and NLoS components, $\mathbf{H}_{\mathrm{NLoS}}[k]$ follows a Rayleigh distribution, and $\mathbf{H}_{\mathrm{LoS}}[k]$ is a deterministic matrix representing the dominant propagation path. Fig. \ref{fr2} presents numerical SER results for $T=4$, $M=N=2$, and $L=16$ for different values of $K_{\mathrm{r}}$ expressed in dB. Note that Rayleigh fading corresponds to the case $K_{\mathrm{r}}=0$ in linear units, where only the NLoS component is present. To obtain $\mathbf{H}_{\mathrm{LoS}}[k]$, we consider free-space channel propagation with isotropic antennas, where the transmit and receive antennas lie in the same horizontal plane and are separated by $20$ m. The spacing between the two transmit antennas and between the two receive antennas is $2\lambda_{\mathrm{c}} = 4$ cm, with $\lambda_{\mathrm{c}} = f_{\mathrm{c}}/c$ being $c$ the speed of light in vacuum, as considered in the presented prototype in Section \ref{PI}. Far-field conditions are assumed, which is also consistent with the use of monopole antennas in the prototype. It can be observed that the system performance improves as $K_{\mathrm{r}}$ increases and the LoS component becomes more dominant. This behavior is expected, since Grassmannian signaling provides MIMO diversity gain, where a single data stream is transmitted through different fading channels to enhance reliability. In this context, Rayleigh fading can be regarded as a worst-case assumption due to its strong fading nature, whereas the presence of a deterministic LoS component ensures a more stable link. However, note that this trend is opposite to that of MIMO systems based on spatial multiplexing, which exploit spatial diversity to create multiple independent virtual channels and transmit multiple data streams simultaneously. For those systems, the presence of a dominant path causes the singular modes of the channel matrix to collapse into one dominant spatial mode, making stream separation ill-conditioned. In such cases, MIMO can provide beamforming for the single stream if CSI is available. Overall, these considerations further support the reliability of the proposed system across different channel conditions, even if this comes at the cost of reduced spectral efficiency in some scenarios compared with the optimal approach for each case.

\section{Prototype Implementation}\label{PI}
In this section, we present the prototype implementation for the system design described in Section \ref{SD}. The experimental evaluation of this prototype is detailed in Section \ref{EE}.

\subsection{General Aspects}\label{GA}

Baseband digital signal processing is developed and implemented with the GNU Radio framework, an open-source platform built on C++ and Python that can run on a general-purpose CPU, such as that of a personal computer. The transmitter and receiver are implemented using Universal Software Radio Peripherals (USRPs), specifically USRP X310 units equipped with two UBX 10–6000 daughterboards each. These devices perform analog-to-digital and digital-to-analog conversion (ADC/DAC), as well as intermediate-frequency conversion. Since USRPs are compatible with GNU Radio, key communication parameters, including gain, carrier frequency, sampling rate, modulation constellation, and power allocation, can be configured via software within the constraints of the hardware. Being the maximum carrier frequency supported by the USRPs devices $6$ GHz, commercial front-end components are added to each device to reach the selected carrier frequency, $f_{\mathrm{c}}= 15$ GHz. In particular, Mini-Circuits ZMDB-44H-K+ mixers, Panda Microwave P/N LAD80180-G24P10\_AC low-noise power amplifiers and waveguide filters  are used, with the intermediate frequency (IF) set to $3$ GHz. Finally, monopole antennas are used at both the transmitter and receiver, with a spacing of $4$ cm, corresponding to $2\lambda_c$ at the selected carrier frequency. This spacing is chosen to enhance diversity under NLoS conditions while preserving a compact design thanks to the shorter wavelength compared with sub-6 GHz bands.

A short packet format is adopted according to the considered potential use cases, as well as for testing purposes. Each packet is composed of three fields: first, an access code of 8 bytes to identify the beginning of the packet; second, a cyclic redundancy check 32 (CRC32) sequence of 4 bytes for error detection; and third, a payload of variable length. The estimation of SER based on the CRC32 field of the received packets is carried out as described in \cite[(14)]{tcom_noma} and \cite{elec}. The sampling rate is set to $5$ Msps per port. Using OFDM with $K=64$ subcarriers, we switch off the center subcarrier to avoid DC interference after IF downconversion and also null $4$ and $3$ subcarriers at the left and right edges of the bandwidth, respectively, to limit out-of-band emissions without further frequency filtering. Thus, $56$ active subcarriers are used for data transmission. Additionally, as detailed in Section \ref{TFS}, a single OFDM synchronization symbol is transmitted for every $32$ data symbols. Since no pilot symbols are transmitted, and with $K_{\mathrm{cp}}=20$, the total data rate is then $\displaystyle\frac{56}{64+20}\times \displaystyle \frac{32} {33} \times(\log_2(L)/T)=0.6465(\log_2(L)/T)$. We consider $L=2$ and $L=16$ with $T=4$, which results in data rates of $0.1616$ and $0.6465$ bps/Hz respectively, or $0.8081$ and $3.2323$ Mbps over the $5$ MHz bandwidth. An entire data packet is transmitted in those $32$ OFDM data symbols between synchronisation symbols, resulting in the transmission of $8$ space-time symbol matrices per subcarrier and packet lengths of $56$ and $224$ bytes for $L=2$ and $L=16$, respectively. On the one hand, note that spectral efficiencies in the kbps/Hz range, although relatively low, are common in URLLC, as well as in wireless technologies such as LoRa, Bluetooth, and Zigbee, and in 5 G control and feedback channels. On the other hand, the FR3 band offers greater spectrum availability than sub-$6$ GHz bands, with the potential to provide considerable data rates even for communication schemes with low spectral efficiency (but high power efficiency).

Finally, note that we consider SER rather than BER, since estimating the latter is more complex and potentially less accurate when both are inferred by evaluating the CRC32 header appended to the packets. However, SER serves as an upper bound on BER, so BER values will always be lower. For the particular family of Grassmannian constellations adopted in this prototype, comparisons between SER and BER measurements with the corresponding bit labeling are shown in \cite{gd}. For single-carrier systems, this family of constellations has the property of requiring only one active transmit antenna per time slot, enabling a potential implementation with a single RF chain. Interestingly, in OFDM, this feature naturally translates into a single active transmit antenna per subcarrier. That is, for each OFDM symbol and a given subcarrier, only one antenna transmits power, since the corresponding complex entries selected from the space-time constellation for all other antennas are zero. Although this necessarily requires $M$ RF chains in the OFDM case, unlike in the single-carrier case, this feature may be of potential use for future integration with sensing applications within the framework of integrated sensing and communications (ISAC). As a further consequence of this feature, the computational complexity of the ML decoder is reduced by a factor of $M$, resulting in $\mathcal{O}(TNL)$, \cite{gd}. Note, nevertheless, that the proposed implementation can support any USTM or Grassmannian constellation design, with decoding computational complexity being the main limiting factor for real-time operation in an OTA system.

\subsection{Time and Frequency Synchronization}\label{TFS}

\begin{figure*}[]
\centering
\begin{tikzpicture}[
    >=Latex,
    cell/.style={draw=black, line width=0.4pt},
    axis/.style={->, thick},
    font=\small
]

\def\nrows{8}
\def\cw{1.2}      
\def\cpw{0.28}    
\def\rh{0.45}
\def\gap{0.75}    
\def\gapb{0.75}   

\pgfmathsetmacro{\xfirstend}{5*\cw + 5*\cpw}
\pgfmathsetmacro{\xsecond}{\xfirstend + \gap}

\pgfmathsetmacro{\xsecondend}{\xsecond + 4*\cw + 4*\cpw}

\pgfmathsetmacro{\xcpfinal}{\xsecondend}
\pgfmathsetmacro{\xsyncfinal}{\xcpfinal + \cpw}
\pgfmathsetmacro{\xend}{\xsyncfinal + \cw + \gapb}

\definecolor{synccolor}{RGB}{180,210,255}
\definecolor{dataone}{RGB}{255,210,160}
\definecolor{datatwo}{RGB}{190,235,190}
\definecolor{datathree}{RGB}{245,190,210}
\definecolor{datafour}{RGB}{220,200,255}
\definecolor{cpcolor}{RGB}{210,210,210}


\fill[cpcolor]   (0,0) rectangle (\cpw,\nrows*\rh);
\fill[synccolor] (\cpw,0) rectangle (\cpw+\cw,\nrows*\rh);

\fill[cpcolor] (\cpw+\cw,0) rectangle (2*\cpw+\cw,\nrows*\rh);
\fill[dataone] (2*\cpw+\cw,0) rectangle (2*\cpw+2*\cw,\nrows*\rh);

\fill[cpcolor] (2*\cpw+2*\cw,0) rectangle (3*\cpw+2*\cw,\nrows*\rh);
\fill[datatwo] (3*\cpw+2*\cw,0) rectangle (3*\cpw+3*\cw,\nrows*\rh);

\fill[cpcolor] (3*\cpw+3*\cw,0) rectangle (4*\cpw+3*\cw,\nrows*\rh);
\fill[datathree] (4*\cpw+3*\cw,0) rectangle (4*\cpw+4*\cw,\nrows*\rh);

\fill[cpcolor] (4*\cpw+4*\cw,0) rectangle (5*\cpw+4*\cw,\nrows*\rh);
\fill[datafour] (5*\cpw+4*\cw,0) rectangle (5*\cpw+5*\cw,\nrows*\rh);

\draw[cell] (0,0) -- (0,\nrows*\rh);
\draw[cell] (\cpw,0) -- (\cpw,\nrows*\rh);
\draw[cell] (\cpw+\cw,0) -- (\cpw+\cw,\nrows*\rh);
\draw[cell] (2*\cpw+\cw,0) -- (2*\cpw+\cw,\nrows*\rh);
\draw[cell] (2*\cpw+2*\cw,0) -- (2*\cpw+2*\cw,\nrows*\rh);
\draw[cell] (3*\cpw+2*\cw,0) -- (3*\cpw+2*\cw,\nrows*\rh);
\draw[cell] (3*\cpw+3*\cw,0) -- (3*\cpw+3*\cw,\nrows*\rh);
\draw[cell] (4*\cpw+3*\cw,0) -- (4*\cpw+3*\cw,\nrows*\rh);
\draw[cell] (4*\cpw+4*\cw,0) -- (4*\cpw+4*\cw,\nrows*\rh);
\draw[cell] (5*\cpw+4*\cw,0) -- (5*\cpw+4*\cw,\nrows*\rh);
\draw[cell] (\xfirstend,0) -- (\xfirstend,\nrows*\rh);

\foreach \j in {0,...,8} {
    \draw[cell] (0,\j*\rh) -- (\xfirstend,\j*\rh);
}

\draw[thick] (0,0) rectangle (\xfirstend,\nrows*\rh);

\node at (\xfirstend + 0.5*\gap,\nrows*\rh/2) {\Huge$\cdot \cdot$};


\fill[cpcolor] (\xsecond,0) rectangle (\xsecond+\cpw,\nrows*\rh);
\fill[dataone] (\xsecond+\cpw,0) rectangle (\xsecond+\cpw+\cw,\nrows*\rh);

\fill[cpcolor] (\xsecond+\cpw+\cw,0) rectangle (\xsecond+2*\cpw+\cw,\nrows*\rh);
\fill[datatwo] (\xsecond+2*\cpw+\cw,0) rectangle (\xsecond+2*\cpw+2*\cw,\nrows*\rh);

\fill[cpcolor] (\xsecond+2*\cpw+2*\cw,0) rectangle (\xsecond+3*\cpw+2*\cw,\nrows*\rh);
\fill[datathree] (\xsecond+3*\cpw+2*\cw,0) rectangle (\xsecond+3*\cpw+3*\cw,\nrows*\rh);

\fill[cpcolor] (\xsecond+3*\cpw+3*\cw,0) rectangle (\xsecond+4*\cpw+3*\cw,\nrows*\rh);
\fill[datafour] (\xsecond+4*\cpw+3*\cw,0) rectangle (\xsecond+4*\cpw+4*\cw,\nrows*\rh);

\draw[cell] (\xsecond,0) -- (\xsecond,\nrows*\rh);
\draw[cell] (\xsecond+\cpw,0) -- (\xsecond+\cpw,\nrows*\rh);
\draw[cell] (\xsecond+\cpw+\cw,0) -- (\xsecond+\cpw+\cw,\nrows*\rh);
\draw[cell] (\xsecond+2*\cpw+\cw,0) -- (\xsecond+2*\cpw+\cw,\nrows*\rh);
\draw[cell] (\xsecond+2*\cpw+2*\cw,0) -- (\xsecond+2*\cpw+2*\cw,\nrows*\rh);
\draw[cell] (\xsecond+3*\cpw+2*\cw,0) -- (\xsecond+3*\cpw+2*\cw,\nrows*\rh);
\draw[cell] (\xsecond+3*\cpw+3*\cw,0) -- (\xsecond+3*\cpw+3*\cw,\nrows*\rh);
\draw[cell] (\xsecond+4*\cpw+3*\cw,0) -- (\xsecond+4*\cpw+3*\cw,\nrows*\rh);
\draw[cell] (\xsecondend,0) -- (\xsecondend,\nrows*\rh);

\foreach \j in {0,...,8} {
    \draw[cell] (\xsecond,\j*\rh) -- (\xsecondend,\j*\rh);
}

\draw[thick] (\xsecond,0) rectangle (\xsecondend,\nrows*\rh);


\fill[cpcolor] (\xcpfinal,0) rectangle (\xcpfinal+\cpw,\nrows*\rh);
\fill[synccolor] (\xsyncfinal,0) rectangle (\xsyncfinal+\cw,\nrows*\rh);

\draw[cell] (\xcpfinal,0) -- (\xcpfinal,\nrows*\rh);
\draw[cell] (\xsyncfinal,0) -- (\xsyncfinal,\nrows*\rh);
\draw[cell] (\xsyncfinal+\cw,0) -- (\xsyncfinal+\cw,\nrows*\rh);

\foreach \j in {0,...,8} {
    \draw[cell] (\xcpfinal,\j*\rh) -- (\xsyncfinal+\cw,\j*\rh);
}

\draw[thick] (\xcpfinal,0) rectangle (\xsyncfinal+\cw,\nrows*\rh);

\node at (\xsyncfinal+\cw+0.5*\gapb,\nrows*\rh/2) {\Huge$\cdot \cdot$};


\draw[axis] (-0.35,0) -- (\xend+0.4,0) node[right] {$t$};
\draw[axis] (0,-1.15) -- (0,\nrows*\rh+1.35) node[above] {$f$};


\draw[<->, line width=1pt] (0,-0.48) -- (\cpw+\cw,-0.48);
\node[above] at ({0.5*(\cpw+\cw)},-0.48) {$t_{\mathrm{OFDM}}$};
\node[below] at ({0.5*(\cpw+\cw)},-0.48) {$16.8\,\mu \textrm{s}$};

\draw[<->, line width=1pt] (\cpw+\cw,-0.48) -- (\xfirstend,-0.48);
\node[above] at ({0.5*((\cpw+\cw)+\xfirstend)},-0.48) {$T$};
\node[below] at ({0.5*((\cpw+\cw)+\xfirstend)},-0.48) {$67.2\,\mu\mathrm{s}$};

\draw[<->, line width=1pt] (0,-1) -- (\xsecondend,-1);
\node[above] at ({0.68*\xsecondend},-1) {$T_{\mathrm{t}}=554.4\,\mu\textrm{s}$};


\draw[<->, line width=1pt] (-0.55,0) -- (-0.55,\nrows*\rh);
\node[rotate=90] at (-0.78,{0.5*\nrows*\rh}) {$5\,\textrm{MHz}$};


\node[align=center] at (\cpw+0.5*\cw,\nrows*\rh+0.45) {Sync.\\Sym.};
\node[align=center] at (\xsyncfinal+0.5*\cw,\nrows*\rh+0.45) {Sync.\\Sym.};

\node[align=center] at (2*\cpw+1.5*\cw,\nrows*\rh+0.45)
    {Sym.~1\\$T^{(1)}$};

\node[align=center] at (3*\cpw+2.5*\cw,\nrows*\rh+0.45)
    {Sym.~2\\$T^{(2)}$};

\node[align=center] at (4*\cpw+3.5*\cw,\nrows*\rh+0.45)
    {Sym.~3\\$T^{(3)}$};

\node[align=center] at (5*\cpw+4.5*\cw,\nrows*\rh+0.45)
    {Sym.~4\\$T^{(4)}$};

\node[align=center] at (\xsecond+\cpw+0.5*\cw,\nrows*\rh+0.45)
    {Sym.~29\\$T^{(1)}$};

\node[align=center] at (\xsecond+\cw+2*\cpw+0.5*\cw,\nrows*\rh+0.45)
    {Sym.~30\\$T^{(2)}$};

\node[align=center] at (\xsecond+2*\cw+3*\cpw+0.5*\cw,\nrows*\rh+0.45)
    {Sym.~31\\$T^{(3)}$};

\node[align=center] at (\xsecond+3*\cw+4*\cpw+0.5*\cw,\nrows*\rh+0.45)
    {Sym.~32\\$T^{(4)}$};


\draw[fill=cpcolor, draw=black, line width=0.4pt]
    (\xsyncfinal+\cw+0.05,\nrows*\rh+0.70) rectangle
    (\xsyncfinal+\cw+0.55,\nrows*\rh+1.15);
\node[anchor=west] at (\xsyncfinal+\cw+0.65,\nrows*\rh+0.925) {\normalsize CP};

\end{tikzpicture}

\caption{Time-frequency system diagram. Sync. and Sym. denote synchronization and symbol, respectively, whereas $T^{(n)}$ denotes the $n$-th time slot within each space-time matrix, with $n=1,2,3,4$. A complete data packet spans $32$ OFDM symbols and $8$ space-time matrices per active subcarrier $\mathbf{X}[k] \in \mathbb{C}^{4 \times 2}$.}
\label{tfdiag}
\end{figure*}
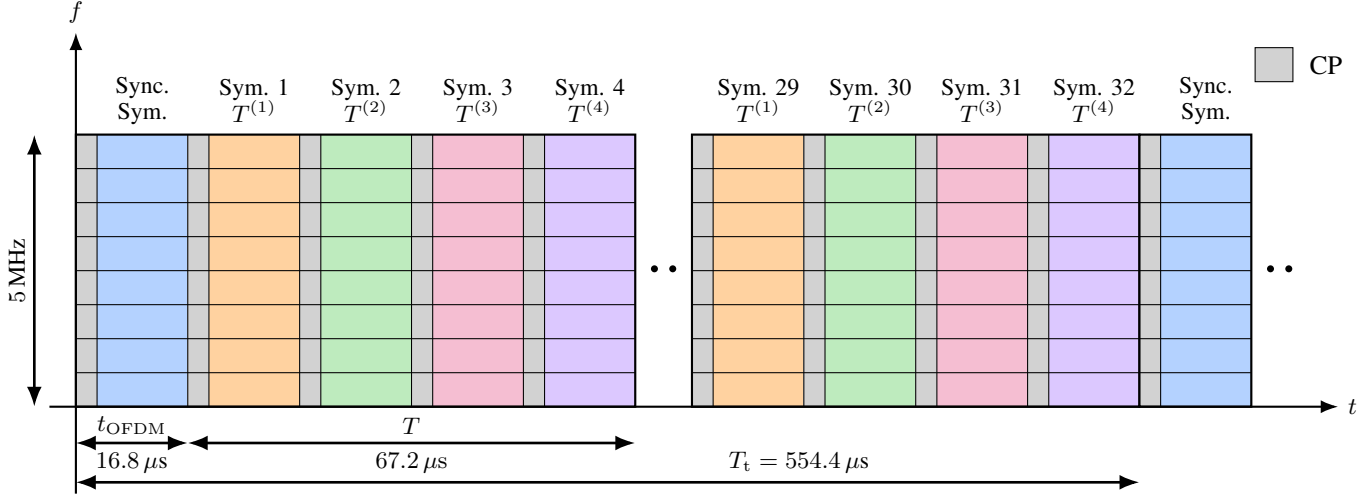

Fig.~\ref{tfdiag} shows the time-frequency system diagram for the transmission of a complete data packet spanning $32$ OFDM symbols, resulting in the transmission of $8$ space-time matrices per subcarrier, $\mathbf{X}[k] \in \mathbb{C}^{4 \times 2}$. Before each packet, an OFDM synchronization symbol is transmitted following the simple yet effective design proposed in \cite{scox}, in which the first half of the synchronization symbol is identical to its second half in the time domain. This structure enables correlation-based frame detection and symbol-timing synchronization at the receiver, as well as CFO estimation and correction. Note that we use a single synchronization symbol rather than two, as also proposed in \cite{scox}, since the expected frequency offsets are much smaller than the subcarrier spacing $\Delta f = 78.125$ kHz (Section~\ref{EC}). As discussed in Section \ref{EC}, the combination of different Doppler shifts and CFO effects caused by hardware imperfections may lead to residual frequency and phase mismatches after applying a single frequency-shift correction, as the one adopted in this scheme. However, as studied in Fig. \ref{ir1}, when the magnitude of each mismatch remains within a limited range, the proposed modulation exhibits only limited performance degradation, even when this results in  coherence assumption within a time block being violated. For the presented prototype, we use an OctoClock CDA-2990 device to provide the USRPs with common time and frequency reference signals, thereby greatly reducing CFO caused by hardware imperfections. Note that these effects are independent of, although they may interact with, Doppler shifts, as they arise from hardware impairments such as I/Q imbalance and local-oscillator drift, whose nature and correction mechanisms are inherently different from those associated with Doppler effects. If the transmitter and receiver are located farther apart, common references can still be provided using, for instance, GPS-disciplined oscillators at both ends. If low-cost hardware constraints prevent the use of such references, it is also possible, as noted in Section \ref{EC}, to use specific Grassmannian constellations designed to mitigate the effects of CFO \cite{cfo}. 

Importantly, the synchronization procedure described above does not provide knowledge of the channel matrices per subcarrier, $\mathbf{H}[k]\in \mathbb{C}^{2\times2}$, and no pilot symbols are transmitted. Although the synchronization symbol could be used for an initial channel estimate, with $M=2$ transmit antennas, at least two orthogonal symbols would need to be introduced to enable the receiver to estimate each independent channel coefficient \cite{mimo_tran}. Furthermore, in such a case, channel variations should be tracked at the receiver through the insertion of pilot symbols within the packet, with the required pilot density depending on the degree of channel variation, as described in Section~ \ref{SD}. On the other hand, in the presented noncoherent prototype, once the start of a packet is detected and frame synchronization is performed, decoding only requires the unknown entries of $\mathbf{H}[k]$ to remain quasi-static over each block of duration $T=67.2$ $\mu$s (Fig. \ref{tfdiag}), even though channel variations over the duration of the packet are expected \cite{gt2}. Regarding the parameters affecting synchronization, e.g., Doppler shifts, CFO effects, and clock mismatches, they only need to remain quasi-constant over the packet data duration of $T_{\mathrm{t}}=554.4$ $\mu$s, after which a new synchronization symbol is transmitted. Moreover, the packet duration can be readily adjusted by varying the number of OFDM symbols to meet this constraint, provided that it remains an integer multiple of the block length $T=4$ OFDM symbols.

\subsection{Transmitter}

\begin{figure*}[t]
\centering
\noindent
\begin{minipage}[t]{0.48\textwidth}
\centering
\resizebox{\linewidth}{!}{%
\begin{tikzpicture}

\node[draw, rectangle, align=center, fill=blue!30,
      minimum height=3.5cm, minimum width=1.0cm]
      (block1) {\rotatebox{90}{Packet Data}};
\node[draw, rectangle, align=center, right=0.5cm of block1, fill=blue!30,
      minimum height=3.5cm, minimum width=1.0cm]
      (block2) {\rotatebox{90}{Source Coding $(\mathcal{X})$}};

\node[draw, rectangle, align=center, right=0.5cm of block2, yshift=1cm,
      fill=blue!30] (block3a) {OFDM\\Processing \& Sync.};
\node[draw, rectangle, align=center, right=0.5cm of block2, yshift=-1cm,
      fill=blue!30] (block3b) {OFDM\\Processing \& Sync.};

\node[draw, rectangle, align=center, right=0.5cm of block3a,
      fill=red!30] (block51) {USRP Tx\\ Port 1};
\node[draw, rectangle, align=center, right=0.5cm of block51,
      fill=red!30] (block61) {Mixer \\+ Filter \\ + Amplifier};

\node[draw, rectangle, align=center, right=0.5cm of block3b,
      fill=red!30] (block5) {USRP Tx \\ Port 2};
\node[draw, rectangle, align=center, right=0.5cm of block5,
      fill=red!30] (block6) {Mixer \\+ Filter \\ + Amplifier};

\draw[->, line width=0.7mm] (block1) -- (block2);
\draw[->, line width=0.7mm] (block2.east |- block3a) -- (block3a.west);
\draw[->, line width=0.7mm] (block2.east |- block3b) -- (block3b.west);
\draw[->, line width=0.7mm] (block3a) -- (block51);
\draw[->, line width=0.7mm] (block51) -- (block61);
\draw[->, line width=0.7mm] (block3b) -- (block5);
\draw[->, line width=0.7mm] (block5) -- (block6);

\draw[line width=0.8mm] (block61.east) -- ++(0.6,0) -- ++(0,0.6)
      node[midway, right, align=left] {Tx Ant.\\ Port 1};
\node[draw, regular polygon, regular polygon sides=3, shape border rotate=180,
      minimum size=1cm, line width=0.8mm, anchor=south, fill=red!30]
      at ($(block61.east) + (0.6,0.6)$) {};

\draw[line width=0.8mm] (block6.east) -- ++(0.6,0) -- ++(0,0.6)
      node[midway, right, align=left] {Tx Ant. \\ Port 2};
\node[draw, regular polygon, regular polygon sides=3, shape border rotate=180,
      minimum size=1cm, line width=0.8mm, anchor=south, fill=red!30]
      at ($(block6.east) + (0.6,0.6)$) {};

\node[draw, circle, minimum size=1.1cm, line width=0.8mm, fill=red!30]
      (osc) at ($ (block61.south)!0.5!(block6.south) + (0,-1.8) $) {};

\draw[line width=0.5mm]
  plot[smooth] coordinates {
    ($(osc.center)+(-0.28,0.00)$)
    ($(osc.center)+(-0.14,0.14)$)
    ($(osc.center)+( 0.00,0.00)$)
    ($(osc.center)+( 0.14,-0.14)$)
    ($(osc.center)+( 0.28,0.00)$)
  };

\node[below=1mm of osc] {Local Oscillator};

\begin{pgfonlayer}{background}
    \draw[line width=0.7mm] ($(osc.north)+(-0.18,0)$) -- ++(0,2.75);
    \draw[line width=0.7mm] ($(osc.north)+( 0.18,0)$) -- ++(0,1.00);
\end{pgfonlayer}

\end{tikzpicture}%
}
\end{minipage}%
\hfill%
\begin{minipage}[t]{0.48\textwidth}
\centering
\raisebox{0mm}{
\resizebox{\linewidth}{!}{%
\begin{tikzpicture}

\node[draw, rectangle, align=center, fill=red!30] (rblock1) {Mixer};
\node[draw, rectangle, align=center, right=0.5cm of rblock1,
      fill=red!30] (rblock2) {USRP Rx\\ Port 1};
\node[draw, rectangle, align=center, right=0.5cm of rblock2,
      fill=blue!30] (rblock3) {Sync. \& \\ OFDM Processing};

\node[draw, rectangle, align=center, right=0.8cm of rblock3, fill=blue!30,
      yshift=-0.75cm, minimum height=3.5cm, minimum width=1.0cm]
      (rblock4) {\rotatebox{90}{Source Decoding $(\mathcal{X})$}};

\node[draw, rectangle, align=center, right=0.8cm of rblock4, fill=blue!30,
      minimum height=3.5cm, minimum width=1.0cm]
      (rblock5) {\rotatebox{90}{Packet Data}};

\draw[->, line width=0.7mm] (rblock1) -- (rblock2);
\draw[->, line width=0.7mm] (rblock2) -- (rblock3);
\draw[->, line width=0.7mm] (rblock3.east) -- (rblock4.west |- rblock3);

\draw[line width=0.8mm] (rblock1.west) -- ++(-0.6,0) -- ++(0,0.6);
\node[left=0.7cm of rblock1, align=left] {Rx Ant.\\ Port 1};
\node[draw, regular polygon, regular polygon sides=3, shape border rotate=180,
      minimum size=1cm, line width=0.8mm, anchor=south, fill=red!30]
      at ($(rblock1.west) + (-0.6,0.6)$) {};

\node[draw, rectangle, align=center, below=1.5cm of rblock1,
      fill=red!30] (rblock1b) {Mixer};
\node[draw, rectangle, align=center, right=0.5cm of rblock1b,
      fill=red!30] (rblock2b) {USRP Rx\\ Port 2};
\node[draw, rectangle, align=center, right=0.5cm of rblock2b,
      fill=blue!30] (rblock3b) {Sync. \& \\ OFDM Processing};

\draw[->, line width=0.7mm] (rblock1b) -- (rblock2b);
\draw[->, line width=0.7mm] (rblock2b) -- (rblock3b);
\draw[->, line width=0.7mm] (rblock3b.east) -- (rblock4.west |- rblock3b);
\draw[->, line width=0.7mm] (rblock4.east) -- (rblock5.west);

\draw[line width=0.8mm] (rblock1b.west) -- ++(-0.6,0) -- ++(0,0.6);
\node[left=0.7cm of rblock1b, align=left] {Rx Ant.\\ Port 2};
\node[draw, regular polygon, regular polygon sides=3, shape border rotate=180,
      minimum size=1cm, line width=0.8mm, anchor=south, fill=red!30]
      at ($(rblock1b.west) + (-0.6,0.6)$) {};

\node[draw, circle, minimum size=1.1cm, line width=0.8mm, fill=red!30]
      (rosc) at ($(rblock1b.south) + (0,-1.1)$) {};

\node[below=1mm of rosc] {Local Oscillator};

\draw[line width=0.5mm]
  plot[smooth] coordinates {
    ($(rosc.center)+(-0.28,0.00)$)
    ($(rosc.center)+(-0.14,0.14)$)
    ($(rosc.center)+( 0.00,0.00)$)
    ($(rosc.center)+( 0.14,-0.14)$)
    ($(rosc.center)+( 0.28,0.00)$)
  };

\begin{pgfonlayer}{background}
    \draw[line width=0.7mm]
        ($(rosc.north)+(-0.18,0)$) --
        ($(rblock1.center)+(-0.18,0)$);

    \draw[line width=0.7mm]
        ($(rosc.north)+(0.18,0)$) --
        ($(rblock1b.center)+(0.18,0)$);
\end{pgfonlayer}

\end{tikzpicture}%
}}
\end{minipage}

\caption{Transmitter and receiver diagrams. Blue and red colors indicate baseband digital signal processing and hardware blocks, respectively.}
\label{txrxschemes}
\end{figure*}
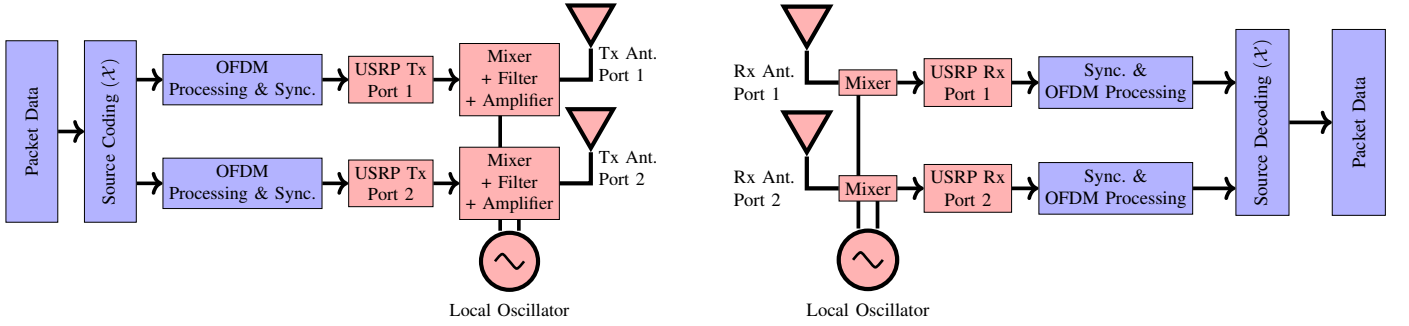

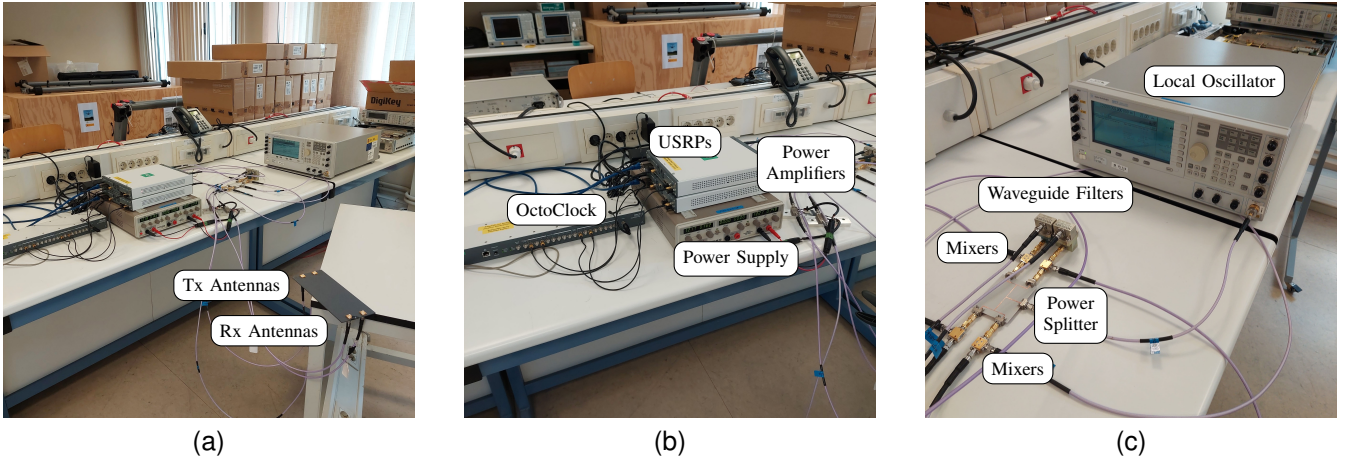
\begin{figure*}
\centering

\subfloat[\label{s1}]{
\begin{tikzpicture}
  \node[inner sep=0] (img) {
    \clipbox{0pt 20pt 0pt 30pt}{
      \includegraphics[angle=-90,origin=c,width=0.3\linewidth]{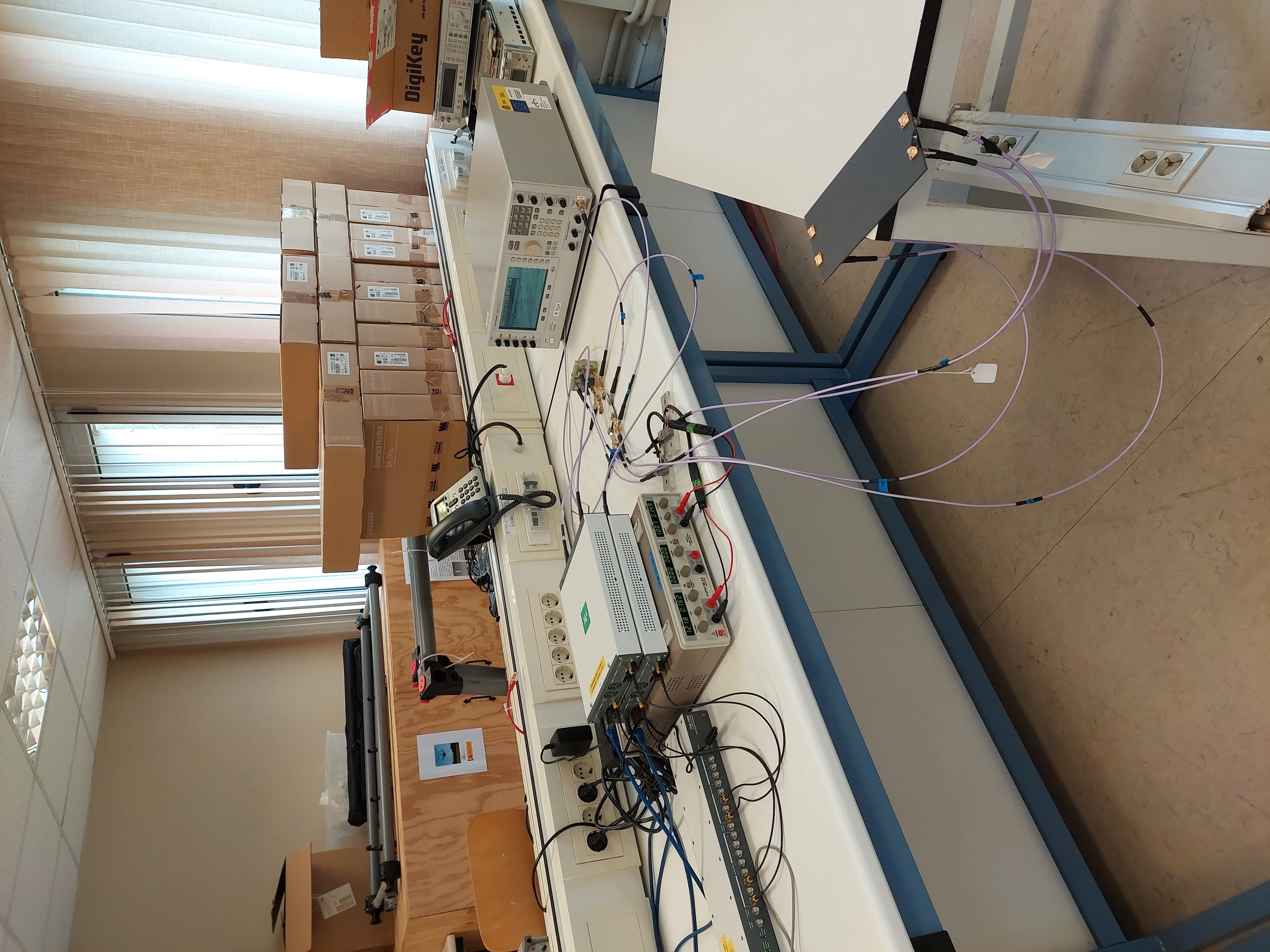}
    }
  };

    \node[draw, fill=white, rounded corners, font=\scriptsize]
    at (0.3, -1) {Tx Antennas};

    \node[draw, fill=white, rounded corners, font=\scriptsize]
    at (0.8, -1.6) {Rx Antennas};

\end{tikzpicture}
}
\hfill
\subfloat[\label{s2}]{
\begin{tikzpicture}
  \node[inner sep=0] (img) {
    \clipbox{0pt 20pt 0pt 30pt}{
      \includegraphics[angle=-90,origin=c,width=0.3\linewidth]{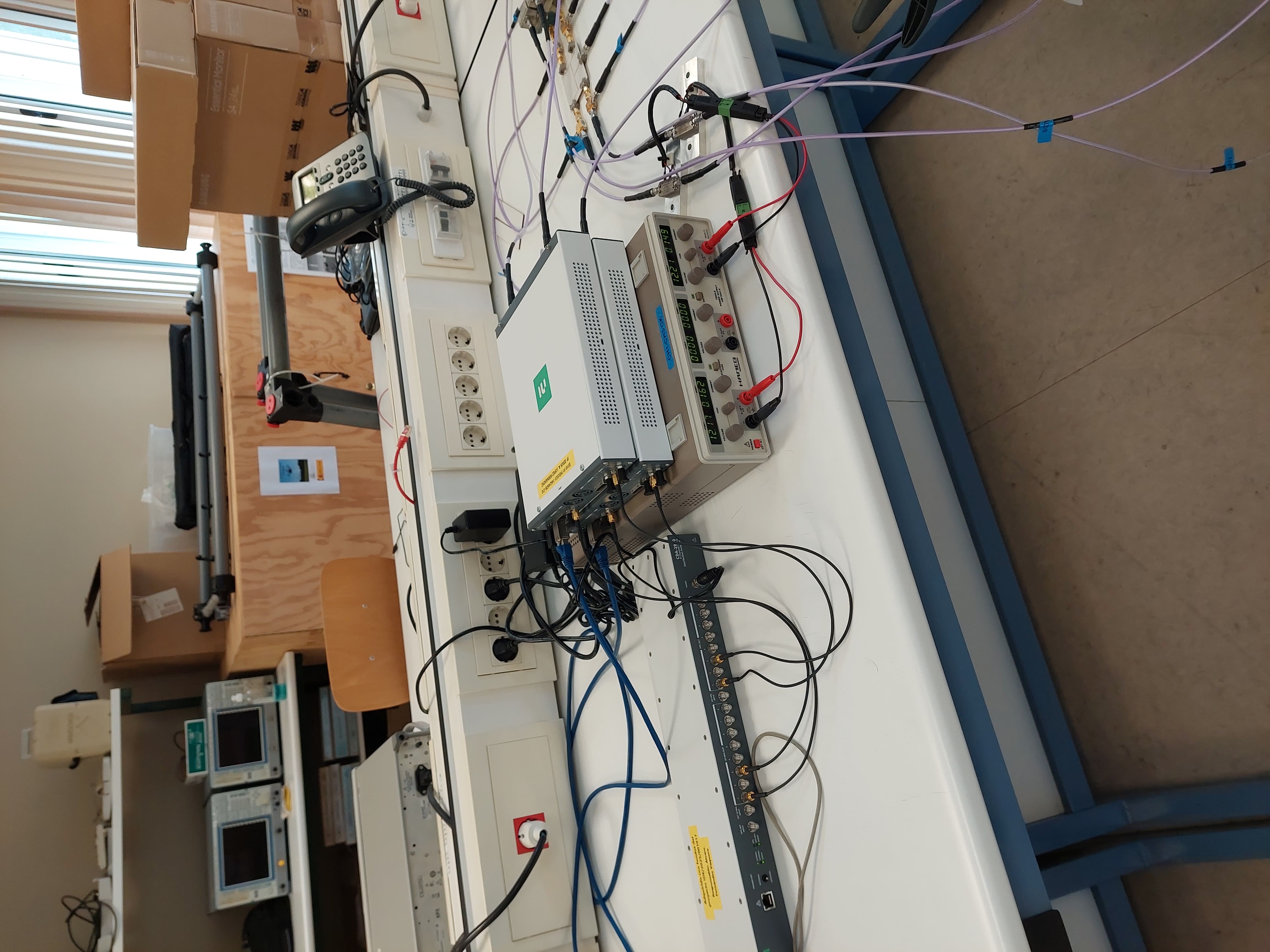}
    }
  };

  \node[
    draw,
    fill=white,
    rounded corners,
    font=\scriptsize,
    align=center
  ] at (0.2, 0.9) {USRPs};

  \node[draw, fill=white, rounded corners, font=\scriptsize]
    at (0.875, -0.65) {Power Supply};

    \node[draw, fill=white, rounded corners, font=\scriptsize, align=center]
    at (1.8, 0.6) {Power \\ Amplifiers};

  \node[
    draw,
    fill=white,
    rounded corners,
    font=\scriptsize,
    align=center
  ] at (-1.5, 0) {OctoClock};

\end{tikzpicture}
}
\hfill
\subfloat[\label{s3}]{
\begin{tikzpicture}
  \node[inner sep=0] (img) {
    \clipbox{0pt 20pt 0pt 30pt}{
      \includegraphics[angle=-90,origin=c,width=0.3\linewidth]{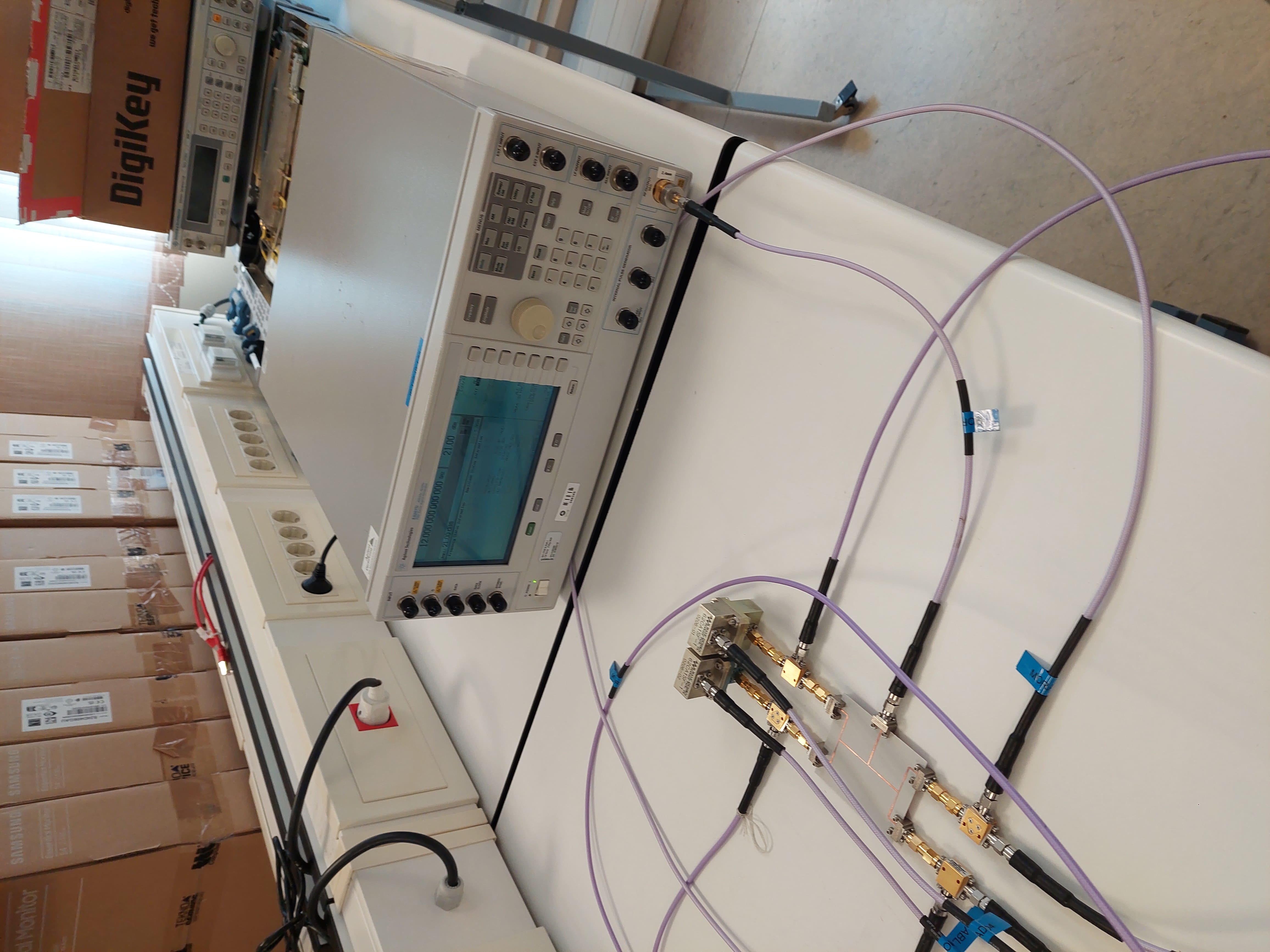}
    }
  };

  \node[draw, fill=white, rounded corners, font=\scriptsize]
    at (1.1, 1.7) {Local Oscillator};

  \node[draw, fill=white, rounded corners, font=\scriptsize]
    at (-1, 0.2) {Waveguide Filters};

  \node[draw, fill=white, rounded corners, font=\scriptsize]
    at (-1.5, -2.1) {Mixers};

    \node[draw, fill=white, rounded corners, font=\scriptsize]
    at (-2.1, -0.5) {Mixers};

  \node[
    draw,
    fill=white,
    rounded corners,
    font=\scriptsize,
    align=center
  ] at (-0.8, -1.4) {Power \\ Splitter};
\end{tikzpicture}
}

\caption{Prototype implementation in the laboratory at the University of Oviedo.
(a) Complete setup. (b) USRPs, OctoClock, power supply and power amplifiers. (c) Mixers, local oscillator, power splitter and waveguide filters.}
\label{setup_photos}
\end{figure*}

Fig. \ref{txrxschemes} shows the transmitter diagram, where blue and red colors indicate baseband digital signal processing and hardware blocks, respectively. Photographs of the hardware setup are shown in Fig. \ref{setup_photos}. First, a packet containing random data is generated according to the structure described in Section \ref{GA}. Its bits are then encoded into space-time matrices using the selected alphabet $\mathcal{X}$ with $L$ constellation points, which determines the data rate. The corresponding matrices are then parallelized across the $K=64$ inputs of the inverse fast Fourier transform (IFFT) in each antenna chain for OFDM processing. Since only $56$ subcarriers are active (Section \ref{GA}), the remaining inactive subcarriers are set to zero. This stage also includes the insertion of the synchronization symbol and the corresponding CPs. The resulting complex baseband samples are passed to the corresponding ports of the USRP used for transmission, which performs DAC and shifts the signals to the IF. The hardware front-end chains then perform frequency upconversion to $f_{\mathrm{c}}=15$ GHz, followed by filtering using WR42 waveguides with a cutoff frequency of $9.488$ GHz to suppress spectral images, and power amplification. The average power given to each antenna is about $-10$ dBm.

\subsection{Receiver}

Fig. \ref{txrxschemes} shows the receiver diagram, the hardware components of which are shown in Fig. \ref{setup_photos}. After frequency downconversion to IF by the mixers, the receiving USRP performs baseband downconversion, low-noise amplification, and ADC. Using the resulting complex digital samples, synchronization is performed as described in Section~\ref{TFS}, followed by OFDM processing for each antenna chain, including CP removal, FFT computation, and serialization of the resulting data. The resulting space-time matrices are then decoded based on the selected alphabet $\mathcal{X}$ using the decoder in (\ref{ml}). Finally, the packet data are rearranged, and the CRC32 header is checked for SER estimation (Section \ref{GA}).

\section{Experimental Evaluation}\label{EE}

In this section, we present the experimental evaluation of the prototype described in Section \ref{PI}. To this end, we compare experimental SER measurements with numerical simulation results under different NLoS and LoS scenarios.

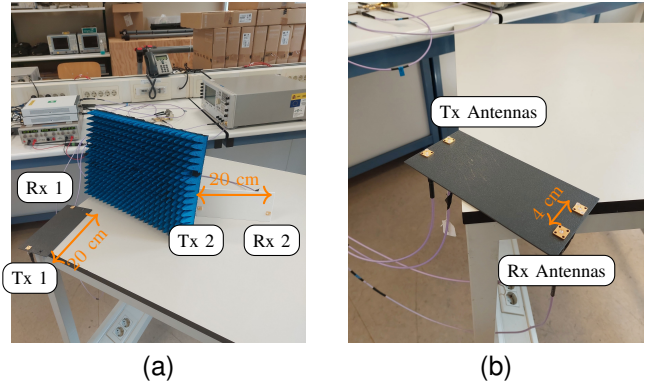
\begin{figure}
\centering

\subfloat[\label{m1}]{
\begin{tikzpicture}
  \node[inner sep=0] (img) {
    \clipbox{0pt 10pt 0pt 10pt}{
      \includegraphics[angle=-90,origin=c,width=0.44\linewidth]{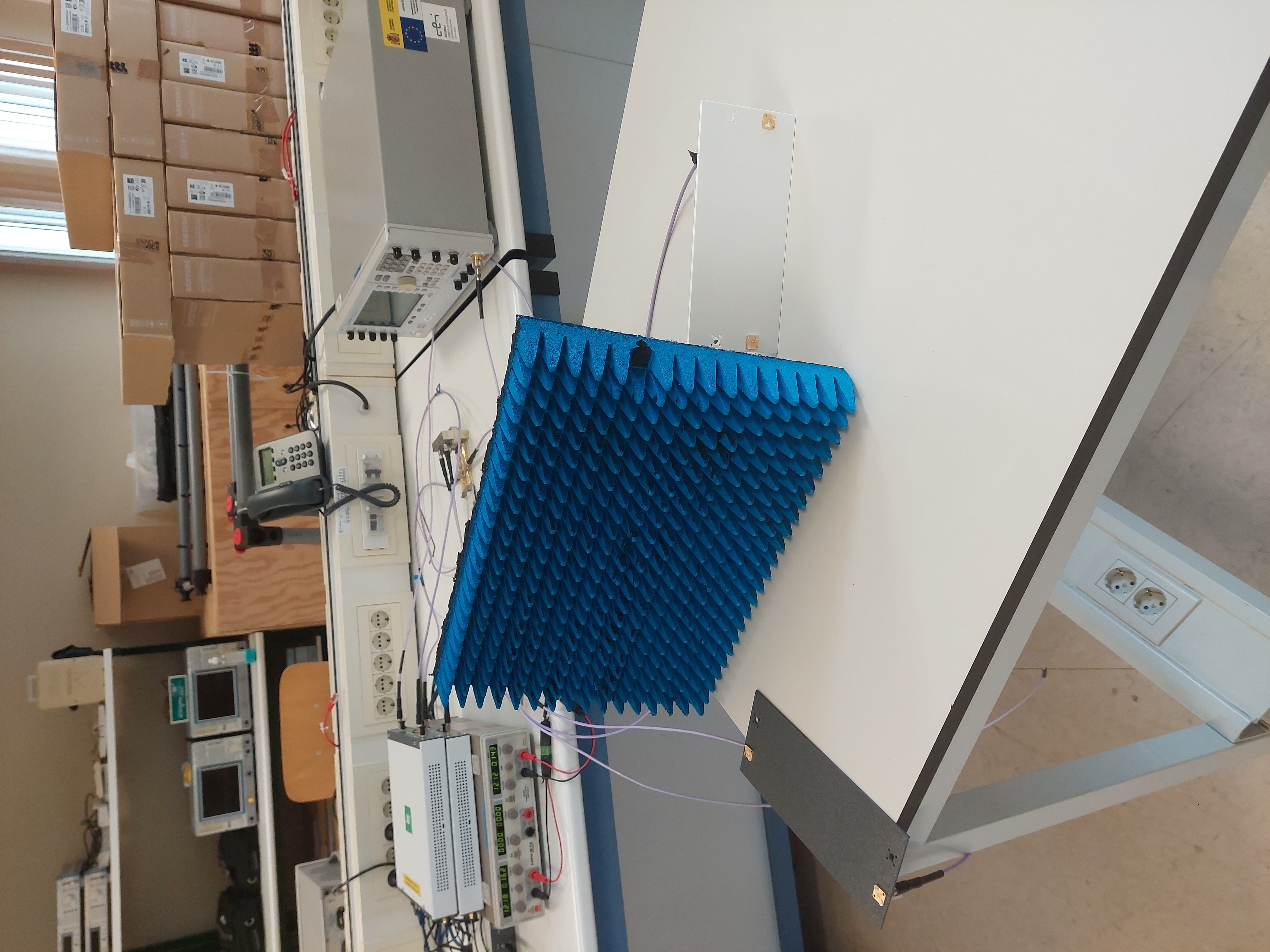}
    }
  };

  \node[draw, fill=white, rounded corners, font=\scriptsize]
    at (-1.5, -0.2) {Rx 1};

\node[draw, fill=white, rounded corners, font=\scriptsize]
    at (-1.7, -1.4) {Tx 1};

\draw[<->, thick, orange]
  (-1.4, -1.2) -- (-0.8, -0.55)
  node[midway, below, sloped, font=\scriptsize, text=orange] {$20$ cm};

  \node[draw, fill=white, rounded corners, font=\scriptsize]
    at (0.5, -0.9) {Tx 2};

    \node[draw, fill=white, rounded corners, font=\scriptsize]
    at (1.5, -0.9) {Rx 2};

\draw[<->, thick, orange]
  (0.5, -0.3) -- (1.5, -0.3)
  node[midway, above, sloped, font=\scriptsize, text=orange] {$20$ cm};
    
\end{tikzpicture}
}
\hfill
\subfloat[\label{m2}]{
\begin{tikzpicture}
  \node[inner sep=0] (img) {
    \clipbox{0pt 10pt 0pt 10pt}{
      \includegraphics[angle=-90,origin=c,width=0.44\linewidth]{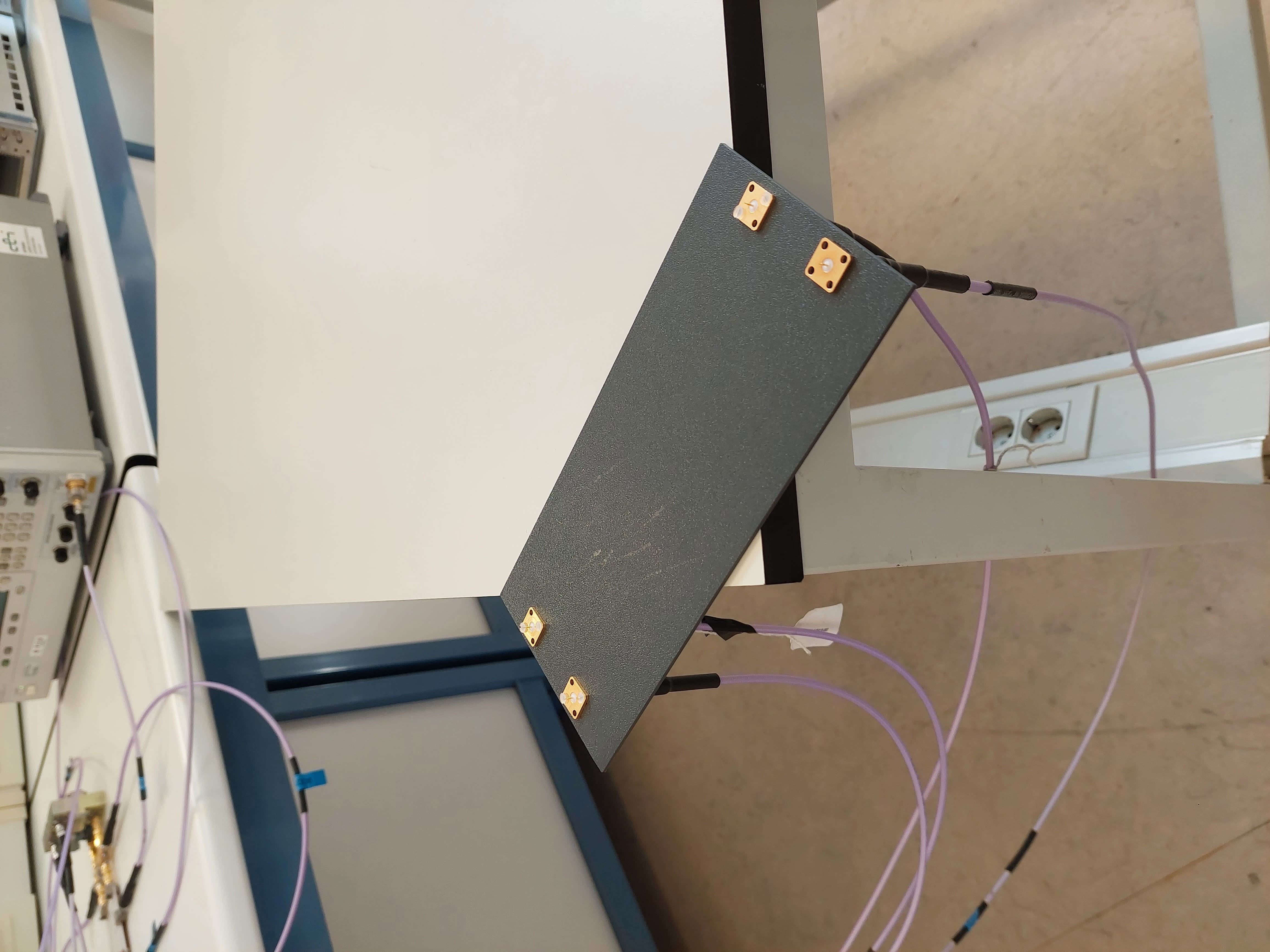}
    }
  };

  \node[draw, fill=white, rounded corners, font=\scriptsize]
    at (-0.1, 0.8) {Tx Antennas};

\draw[<->, thick, orange]
  (0.7, -0.7) -- (0.97, -0.37)
  node[midway, above, sloped, font=\scriptsize, text=orange] {$4$ cm};

  \node[draw, fill=white, rounded corners, font=\scriptsize]
    at (0.8, -1.3) {Rx Antennas};

\end{tikzpicture}
}

\caption{Measurement setups. (a) Rayleigh block-fading NLoS scenario. (b) LoS Scenario.}
\label{meas_photos}
\end{figure}

\subsection{NLoS Measurements}

First, we aim to reproduce the most standard channel model for the study of noncoherent communications, namely Rayleigh block fading. Under this assumption, the entries of the channel matrix for each subcarrier are i.i.d. and follow a zero-mean, unit-variance complex Gaussian distribution, i.e., $h_{kij} \sim \mathcal{CN}(0,1)$, where $h_{kij}$ denotes the channel coefficient in the $i$-th row and $j$-th column of $\mathbf{H}[k]$. Each entry remains constant over $T$ symbols and then changes to an independent realization in the next block. Although this is an ideal model, it provides a useful abstraction for capturing the dispersive behavior of rich-scattering environments in the absence of a dominant propagation path, while retaining a simple statistical characterization and analytical tractability. Consequently, it is widely adopted for constellation design and performance evaluation in noncoherent systems \cite{nonch_review}.

To obtain reproducible SER measurements that mimic the desired channel model while maintaining static antenna positions, we introduce a controlled channel distortion and artificial noise (AN) in each chain of the transmitter diagram in Fig. \ref{txrxschemes}, which are applied to the digital samples before they are passed to the USRP. Introducing those channel effects at the transmitter ensures that they are reflected in the analog OTA radiated signal, which is essential for evaluating the prototype beyond digital simulations. To reproduce Rayleigh block fading by acting only at the transmitter, we adopt the following approach. First, we physically decouple the transmit and receive antennas into two independent wireless links, as shown in Fig. \ref{m1}. This is achieved by rotating one antenna pair by $90^\circ$ with respect to the other to exploit polarization discrimination, since the monopoles are linearly polarized, and by placing a piece of absorber foam between the two pairs. This results in an approximately diagonal physical channel matrix, i.e., $\mathbf{H}[k] \approx \operatorname{diag}(h_{k11},h_{k22})$, since $h_{k12},h_{k21} \approx 0$. Second, we multiply the digital samples by a matrix $\mathbf{H}_{\mathrm{c}}[k]\in \mathbb{C}^{2\times2}$ whose entries are i.i.d and are denoted as $h_{kji}^{\mathrm{c}}$, $h_{kji}^{\mathrm{c}} \sim \mathcal{CN}(0,1)$, and add an AN matrix $\sqrt{1/\rho}\mathbf{W}_{\mathrm{c}}[k]$, $\mathbf{W}_{\mathrm{c}}[k]\in \mathbb{C}^{2\times2}$, where $w_{kij}^{\mathrm{c}} \sim \mathcal{CN}(0,1)$. This results in a received matrix

\begin{equation}\label{Yofdm_nlos1}
\mathbf{Y}[k] =\left(\sqrt{\frac{T}{M}} \mathbf{X}[k]\mathbf{H}_{\mathrm{c}}[k] + \sqrt{1/\rho}\mathbf{W}_{\mathrm{c}}[k]\right)\mathbf{H}[k] + \mathbf{Z}[k],
\end{equation} where $\mathbf{Z}[k]$ denotes the physical hardware noise, whose variance can be assumed constant for static transmitter and receiver positions and is denoted by $\sigma_z^2$. If we assume that the variance of this noise is negligible compared with that of the scaled AN, we can rewrite the received matrix as:

\begin{equation}\label{Yofdm_nlos2}
\mathbf{Y}[k] \approx \sqrt{\frac{T}{M}} \mathbf{X}[k]\mathbf{H}_{\mathrm{c}}[k]\mathbf{H}[k] + \sqrt{1/\rho}\mathbf{W}_{\mathrm{c}}[k]\mathbf{H}[k],
\end{equation} moreover,

\begin{equation}\label{HYofdm_nlos2}
\mathbf{H}_{\mathrm{c}}[k]\mathbf{H}[k] = \begin{bmatrix}
h_{k11}^{\mathrm{c}} h_{k11}& h_{k12}^{\mathrm{c}} h_{k22}\\
h_{k21}^{\mathrm{c}} h_{k11} & h_{k22}^{\mathrm{c}} h_{k22}
\end{bmatrix},
\end{equation} where, since identical hardware elements are used for both links, we have $\left|h_{k11}\right| \approx \left|h_{k22}\right|$, with both coefficients remaining constant. Under this assumption, the desired Rayleigh fading behavior is achieved through the coefficients $h_{kij}^{\mathrm{c}}$, while the SNR is still determined by $\rho$, since the power scaling introduced by the physical channel is also applied to $\mathbf{W}_{\mathrm{c}}[k]$.

\begin{figure}
\centering
\includegraphics[width=0.8\linewidth]{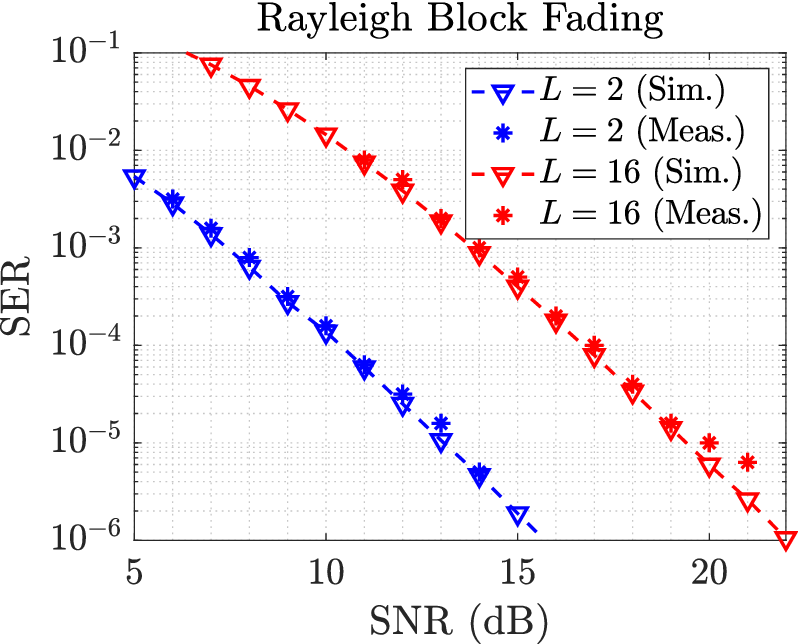}
\caption{SER measurements (Meas.) compared with simulation results (Sim.) for the Rayleigh block fading scenario.}
\label{fr1}
\end{figure}

Fig. \ref{fr1} compares the measured and simulated SER values for constellation sizes $L=2$ and $L=16$. The numerical results are obtained through Monte Carlo simulations using the expression in (\ref{Yofdm}). The good agreement between the simulations and measurements validates the prototype under block-fading conditions. As expected, at higher SNR values (around $20$ dB), the measured SER is slightly higher than the simulated SER because the physical noise introduced by the hardware becomes non-negligible relative to the added AN.

Finally, we also conduct some basic Doppler testing in this scenario by adding a controlled frequency shift noted as $\Delta f_\mathrm{d}^{\mathrm{c}}$ at both transmit chains. Within the maximum design range of $\Delta f_{\mathrm{d}}^{\mathrm{c}} \leq f_{\mathrm{dmax}}=3.5$ kHz (Section~\ref{EC}), this single frequency shift is readily corrected by the frequency synchronization described in Section \ref{TFS} and therefore has no impact on the measured SER. In contrast, values on the same order of magnitude as $\Delta f = 78.125$ kHz, e.g., $\Delta f_{\mathrm{d}}^{\mathrm{c}} = 50$ kHz, cannot be corrected using a single synchronization symbol and result in severe ICI, strongly degrading communication performance. Additionally, note that while this scenario reproduces the fading effects associated with multipath propagation and a dispersive channel, the system design described in Section \ref{EC} would also protect against the effects of delay spread, up to the chosen design value of $\tau_{\mathrm{rms}}=955$ ns.

\subsection{LoS Measurements}

For measurements in the LoS scenario, we position the transmit and receive antenna pairs with the already mentioned spacing of $2\lambda_c=4$ cm, as illustrated in Fig. \ref{m2} and specified in the prototype design in Section \ref{PI}. The transmit and receive antennas lie in the same plane and are separated by $20$ cm. This scenario clearly exhibits a strong dominant path, and although some multipath components are naturally present due to the monopole radiation patterns, which are approximately omnidirectional in the horizontal plane, their effects are expected to be negligible given the carrier frequency of $f_{\mathrm{c}}=15$ GHz and the short separation between the transmit and receive antennas. In this case, only AN is added at the transmitter with a scaling factor of $\sqrt{1/(2\rho)}$, since none of the entries of the physical channel matrix $\mathbf{H}[k]$ are zero and, consequently, the noise components added to each transmit chain contribute to both receive chains. Again, we assume that the variance of the physical hardware noise is negligible compared with that of the AN, which is reasonable given the low-SNR regime evaluated here.

\begin{figure}
\centering
 \subfloat[\normalsize \label{sparam_m}]{%
       \includegraphics[width=0.8\linewidth]{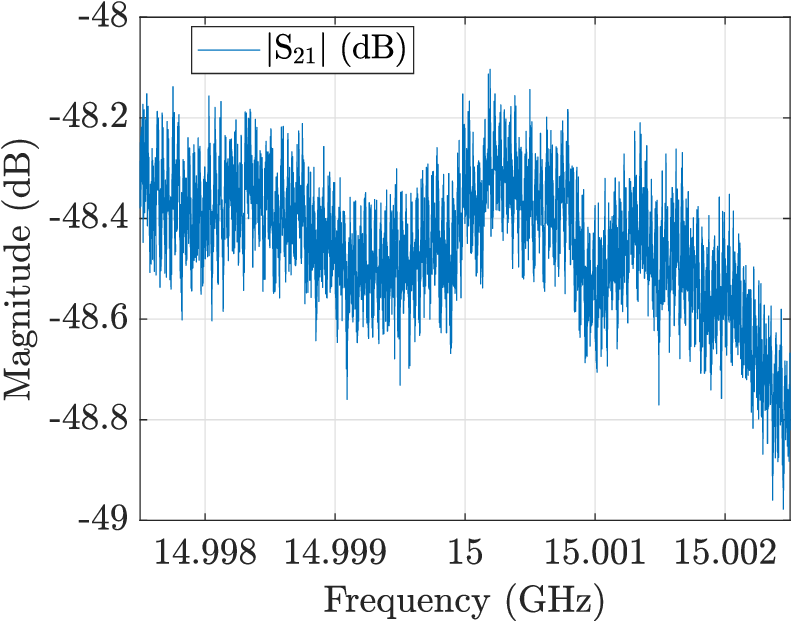}}
  \hfill
  \subfloat[\normalsize \label{sparam_p}]{%
        \includegraphics[width=0.8\linewidth]{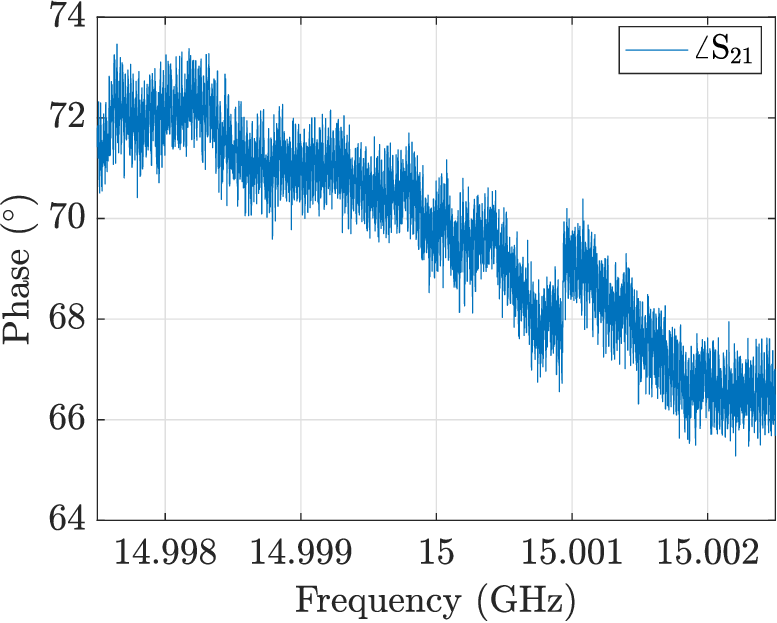}}
  \hfill
  \caption{Measured $|\mathrm{S}_{21}|$ for the LoS Scenario shown in Fig. \ref{m2} between the left transmit antenna and the left receive antenna, from the viewpoint of the transmitter. (a) Magnitude, $|\mathrm{S}_{21}|$. (b) Phase,  $\mathrm{S}_{21}$.}
  \label{s21}
\end{figure}

\begin{figure}
\centering
\includegraphics[width=0.8\linewidth]{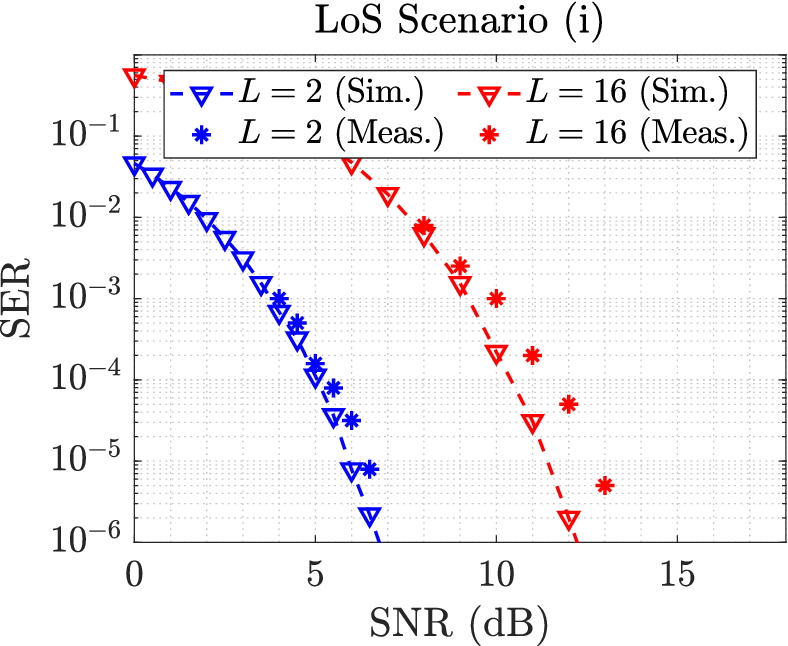}
\caption{SER measurements (Meas.) compared with simulation results (Sim.) for the LoS scenario in Fig. \ref{m2}.}
\label{flos}
\end{figure}

To evaluate the system performance, we first characterize the $\mathrm{S}$ parameters between each antenna link in this scenario with a vector network analyzer (VNA). Fig. \ref{s21} presents the measured $\mathrm{S}_{21}$ between the left transmit antenna and the left receive antenna, from the viewpoint of an observer behind the transmitter. It can be observed that, despite the LoS conditions and the relatively small bandwidth of $5$ MHz, the phase varies by up to $8^\circ$ across the entire bandwidth, which would have an impact on system performance if OFDM were not applied, as modeling the entire channel with a single channel coefficient would then be inaccurate. Fig. \ref{flos} shows SER measurements versus simulation results. For the simulation results, $\mathbf{H}[k]$ is determined by the measured $\mathrm{S}_{21}$ for the four antenna links, averaging for each subcarrier all the obtained points within its bandwidth, and normalizing in magnitude to the absolute maximum across all links and frequencies. Note that, since AN is added at the transmitter, this normalization or channel scaling does not affect the evaluated SNR value. Good agreement is also observed in this scenario for constellation sizes of $L=2$ and $L=16$. For $L=16$, a slight deviation of about $1$ dB is observed as the SNR increases, occurring at lower SNR values than the deviation observed in Fig.~\ref{fr1}. Nonetheless, this behavior is expected to some extent given the abrupt SER slope in this case, which makes any mismatch between the simulations and the hardware setup more apparent in the comparison.

\section{Conclusion}\label{C}

In this work, we present a prototype implementation of a noncoherent MIMO communication system in the envisioned FR3 upper mid-band ($f_{\mathrm{c}} = 15$ GHz) based on Grassmannian signaling. Although this family of noncoherent solutions has been extensively studied theoretically, only a very limited number of experimental works have been reported in sub-6 GHz frequency bands, and, to the best of our knowledge, this is the first Grassmannian or USTM prototype reported in the literature beyond 6 GHz.

We begin by reviewing the fundamentals of Grassmannian signaling combined with OFDM for communication without CSI at either the transmitter or the receiver. Then, with the aim of presenting a working prototype in a research-oriented manner, we consider the specific input parameters $f_{\mathrm{c}} = 15$ GHz and $M=N=2$ transmit and receive antennas, while deliberately adopting general design conditions for the system. In particular, we target a system capable of operating under the larger delay spreads associated with urban areas and the higher velocities associated with rural areas, while reliably supporting both LoS and NLoS conditions without CSI. This versatility is particularly appealing for terrestrial URLLC communications in V2X and V2V scenarios, including UAV-to-UAV and UAV-to-vehicle safety-critical and control links. Based on these specifications, we detail the rationale behind each system design choice and then present the hardware prototype implementation. OTA testing is based on SER measurements under NLoS (Rayleigh block-fading) and LoS (strong dominant-path) conditions for different constellation sizes, showing good agreement with software simulations and thereby validating the prototype implementation.

As discussed throughout the paper, the presented design can support a larger number of antennas without major modifications, with the advantage that phase coherence across antennas is not required owing to the noncoherent nature of Grassmannian signaling. However, ML detection at increasing data rates can result in prohibitive computational complexity, motivating further research on reduced-complexity decoding for USTM and Grassmannian signaling. Several such solutions are already available in the literature, at the cost of some degradation in SER performance. Finally, although in this paper we identify $f_{\mathrm{c}} = 15$ GHz as a possible non-strict upper limit for the practical application of Grassmannian signaling under our working assumptions, the potential use of these or similar noncoherent schemes in satellite communications and at higher carrier frequencies remains an interesting open research question.

The source code and GNU Radio files for the SDR-based prototype implementation will be made available if the paper is accepted for publication.

\bibliographystyle{ieeetr}
\bibliography{references}

@ARTICLE{fr3_nlos,
  author={Shakya, Dipankar and Ying, Mingjun and Rappaport, Theodore S. and Poddar, Hitesh and Ma, Peijie and Wang, Yanbo and Al-Wazani, Idris},
  journal={IEEE Open J. Commun. Soc.}, 
  title={{Comprehensive FR1(C) and FR3 Lower and Upper Mid-Band Propagation and Material Penetration Loss Measurements and Channel Models in Indoor Environment for 5G and 6G}}, 
  year={2024},
  month={July},
  volume={5},
  number={},
  pages={5192-5218},
  doi={10.1109/OJCOMS.2024.3431686}}

@ARTICLE{csi_track,
  author={Zhao, Jianwei and Gao, Feifei and Jia, Weimin and Zhang, Shun and Jin, Shi and Lin, Hai},
  journal={IEEE Trans. Wireless Commun.}, 
  title={{Angle Domain Hybrid Precoding and Channel Tracking for Millimeter Wave Massive MIMO Systems}}, 
  year={2017},
  month = {Aug.},
  volume={16},
  number={10},
  pages={6868-6880},
  doi={10.1109/TWC.2017.2732405}}

@ARTICLE{ses_mag,
  author={Feito, Sergio M. and Pendás-Recondo, {\'A}lvaro and Pino, Marcos R. and Ayestarán, Rafael G. and López-Fernández, Jesús A. and Vaquero, {\'A}lvaro F. and Arrebola, Manuel},
  journal={IEEE Commun. Mag.}, 
  title={{Could Smart Electromagnetic Skins Play a Key Role in mmWave Wireless Communications Deployment?}}, 
  year={2026},
  month={May},
  volume={64},
  number={6},
  pages={81-87},
  doi={10.1109/MCOM.001.2500594}}

@ARTICLE{csitm,
  author={Alkhateeb, Ahmed and El Ayach, Omar and Leus, Geert and Heath, Robert W.},
  journal={IEEE J. Sel. Top. Signal Process.}, 
  title={{Channel Estimation and Hybrid Precoding for Millimeter Wave Cellular Systems}}, 
  year={2014},
  volume={8},
  number={5},
moth={Oct.},
  pages={831-846},
  doi={10.1109/JSTSP.2014.2334278}}

@ARTICLE{index_mod,
  author={Vilà-Insa, Marc and Martí, Aniol and Lamarca, Meritxell and Riba, Jaume},
  journal={IEEE Wireless Commun. Lett.}, 
  title={{Low-Complexity Detection of Permutational Index Modulation for Noncoherent Communications}}, 
  year={2025},
  month={July},
  volume={14},
  number={10},
  pages={3059-3063},
  doi={10.1109/LWC.2025.3585351}}

@ARTICLE{diff_2,
  author={Chen-Hu, Kun and Liu, Yong and Armada, Ana García},
  journal={IEEE Trans. Veh. Technol}, 
  title={{Non-Coherent Massive MIMO-OFDM Down-Link Based on Differential Modulation}}, 
  year={2020},
  month={July},
  volume={69},
  number={10},
  pages={11281-11294},
  doi={10.1109/TVT.2020.3008913}}

@ARTICLE{energy_2,
    author={Bica{\"{\i}}s, Simon and Falempin, Alexis and Dor{\'e}, Jean-Baptiste and Savin, Valentin},
  journal={IEEE Trans. Wireless Commun.}, 
  title={{Design and Analysis of MIMO Systems Using Energy Detectors for Sub-THz Applications}}, 
  year={2022},
  month={Nov.},
  volume={21},
  number={6},
  pages={3678-3690},
  doi={10.1109/TWC.2021.3123220}}

@ARTICLE{diff_survey,
  author={Xu, Chao and Ishikawa, Naoki and Rajashekar, Rakshith and Sugiura, Shinya and Maunder, Robert G. and Wang, Zhaocheng and Yang, Lie-Liang and Hanzo, Lajos},
  journal={IEEE Access}, 
  title={{Sixty Years of Coherent Versus Non-Coherent Tradeoffs and the Road From 5G to Wireless Futures}}, 
  year={2019},
  month={Dec.},
  volume={7},
  number={},
  pages={178246-178299},
  doi={10.1109/ACCESS.2019.2957706}}

@ARTICLE{energy_based,
  author={Vilà-Insa, Marc and Martí, Aniol and Riba, Jaume and Lamarca, Meritxell},
  journal={IEEE Trans. Wireless Commun.}, 
  title={{Quadratic Detection in Noncoherent Massive SIMO Systems Over Correlated Channels}}, 
  year={2024},
  month = {June},
  volume={23},
  number={10},
  pages={14259-14272},
  doi={10.1109/TWC.2024.3411164}}

@ARTICLE{csi_mmwave,
  author={Hassan, Kais and Masarra, Mohammad and Zwingelstein, Marie and Dayoub, Iyad},
  journal={IEEE Open J. Commun. Soc.}, 
  title={{Channel Estimation Techniques for Millimeter-Wave Communication Systems: Achievements and Challenges}}, 
  year={2020},
  month={Aug.},
  volume={1},
  number={},
  pages={1336-1363},
  doi={10.1109/OJCOMS.2020.3015394}}

@ARTICLE{csi1,
  author={Love, David J. and Heath, Robert W. and N. Lau, Vincent K. and Gesbert, David and Rao, Bhaskar D. and Andrews, Matthew},
  journal={IEEE J. Sel. Areas Commun.}, 
  title={{An overview of limited feedback in wireless communication systems}}, 
  year={2008},
  month = {Oct.},
  volume={26},
  number={8},
  pages={1341-1365},
  doi={10.1109/JSAC.2008.081002}}

@ARTICLE{csi2,
  author={Elijah, Olakunle and Leow, Chee Yen and Rahman, Tharek Abdul and Nunoo, Solomon and Iliya, Solomon Zakwoi},
  journal={IEEE Commun. Surveys Tuts.}, 
  title={{A Comprehensive Survey of Pilot Contamination in Massive MIMO—5G System}}, 
  year={2016},
  month = {Nov.},
  volume={18},
  number={2},
  pages={905-923},
  doi={10.1109/COMST.2015.2504379}}

@INPROCEEDINGS{exp1,
  author={Zhu, W. and Liu, D. and Browne, D. and Fitz, M. P.},
  booktitle={Conference Record of the Thirty-Eighth Asilomar Conference on Signals, Systems and Computers, 2004},
  title={{Experiments in Space-Time Modulation}},
  year={2004},
  volume={1},
  number={},
  pages={782-786},
  address={Pacific Grove, CA, USA},
  doi={10.1109/ACSSC.2004.1399242}
}

@INPROCEEDINGS{exp2,
  author={Fanjul, Jacobo and Ibáñez, Jesús and Santamaria, Ignacio and Loucera, Carlos},
  booktitle={Ad-hoc, Mobile, and Wireless Networks: 16th International Conference on Ad Hoc Networks and Wireless, ADHOC-NOW 2017},
  title={{Experimental Evaluation of Non-coherent MIMO Grassmannian Signaling Schemes}},
  year={2017},
  volume={},
  number={},
  pages={221-230},
  address={Messina, Italy},
  doi={10.1007/978-3-319-67910-5_18}
}

@phdthesis{exp3,
      author = {Cuevas, Diego},
      title = {{Advanced {Grassmannian} Constellation Designs for Noncoherent {MIMO} Communications}},
      school = {Universidad de Cantabria, Spain},
      year = {2024},
      month = nov,
      note = {Software available at \url{https://github.com/diegocuevasfdez/grassbox/}}
    }

@INPROCEEDINGS{exp4,
  author={Cuevas, Diego and Gutiérrez, Mikel and Ibáñez, Jesús and Santamaria, Ignacio},
  booktitle={2025 59th Asilomar Conference on Signals, Systems, and Computers},
  title={{Low Probability of Detection Communication Using Noncoherent Grassmannian Signaling}},
  year={2025},
  volume={},
  number={},
  pages={514-518},
  address={Pacific Grove, CA, USA},
  doi={10.1109/IEEECONF67917.2025.11443893}
}

@ARTICLE{tcom_noma,
  author={Pendás-Recondo, {\'A}lvaro and Feito, Sergio M. and Vaquero, {\'A}lvaro F. and López-Fernández, Jesús Alberto and Ayestarán, Rafael González and Pino, Marcos R. and Arrebola, Manuel},
  journal={IEEE Trans. Commun.}, 
  title={{A Practical Implementation of mmWave NOMA with IRS-aided Indoor Coverage Extension}}, 
  year={2025},
month ={Dec.},
  volume={73},
  number={12},
  pages={13741-13757},
  doi={10.1109/TCOMM.2025.3610214}}

@Article{elec,
AUTHOR = {Pendás-Recondo, Alvaro and López-Fernández, Jesús Alberto and González-Ayestarán, Rafael},
TITLE = {{Software-Defined Radio Testbed for I/Q Imbalanced Single-Carrier Communication Systems}},
JOURNAL = {Electronics},
VOLUME = {13},
YEAR = {2024},
month = {July},
NUMBER = {15},
ARTICLE-NUMBER = {3002},
URL = {https://www.mdpi.com/2079-9292/13/15/3002},
ISSN = {2079-9292},
DOI = {10.3390/electronics13153002}
}

@INPROCEEDINGS{cfo,
  author={Cuevas, Diego and Álvarez-Vizoso, Javier and Gutiérrez, Mikel and Santamaria, Ignacio and Tuček, Vít and Peters, Gunnar},
  booktitle={ICASSP 2024 - 2024 IEEE International Conference on Acoustics, Speech and Signal Processing (ICASSP)}, 
  title={{Hardware Impairments-Aware Design of noncoherent Grassmannian Constellations}}, 
  year={2024},
  volume={},
  number={},
  pages={9086-9090},
  address = {Seoul, Korea Republic},
  doi={10.1109/ICASSP48485.2024.10447367}}

@ARTICLE{gd,
  author={Pendás-Recondo, {\'A}lvaro and Pendás-Recondo, Enrique},
  journal={IEEE Trans. Wireless Commun.}, 
  title={{A Structured Family of Grassmannian Constellations via Geodesic Mapping for MIMO Noncoherent Communications}}, 
  year={2026},
  month = {Aug.},
  volume={25},
  number={},
  pages={21089-21104},
  doi={10.1109/TWC.2026.3717208}}

@ARTICLE{cubesplit,
  author={Ngo, Khac-Hoang and Decurninge, Alexis and Guillaud, Maxime and Yang, Sheng},
  journal={IEEE Trans. Wireless Commun.}, 
  title={{Cube-Split: A Structured Grassmannian Constellation for Non-Coherent SIMO Communications}}, 
  year={2020},
month={Mar.},
  volume={19},
  number={3},
  pages={1948-1964},
  doi={10.1109/TWC.2019.2959781}}

@ARTICLE{joint_geo,
  author={Ngo, Khac-Hoang and Yang, Sheng and Guillaud, Maxime and Decurninge, Alexis},
  journal={IEEE Trans. Inf. Theory}, 
  title={{Joint Constellation Design for Noncoherent MIMO Multiple-Access Channels}}, 
  year={2022},
month={July},
  volume={68},
  number={11},
  pages={7281-7305},
  doi={10.1109/TIT.2022.3189254}}

@ARTICLE{nonch_review,
  author={Ngo, Khac-Hoang and Cuevas, Diego and Gil, Ruben De Miguel and Baeza, Victor Monzon and Armada, Ana Garcia and Santamaria, Ignacio},
  journal={IEEE Open J. Veh. Technol.}, 
  title={{Noncoherent MIMO Communications: Theoretical Foundation, Design Approaches, and Future Challenges}}, 
  year={2025},
month ={(early access)},
  volume={},
  number={},
  pages={1-20},
  doi={10.1109/OJVT.2025.3649122}}

@ARTICLE{simo_grass,
  author={Cuevas, Diego and Álvarez-Vizoso, Javier and Beltrán, Carlos and Santamaria, Ignacio and Tuček, Vít and Peters, Gunnar},
  journal={IEEE Trans. Wireless Commun. }, 
  title={{Constellations on the Sphere With Efficient Encoding-Decoding for Noncoherent Communications}}, 
  year={2024},
month={Mar.},
  volume={23},
  number={3},
  pages={1886-1898},
  doi={10.1109/TWC.2023.3292935}}

@ARTICLE{simo_nograss1,
  author={Meng, Huan and Zheng, Lin and Yang, Chao and Chen, Jianmei and Deng, Xiaofang and Wang, Junyi},
  journal={IEEE Trans. Commun.}, 
  title={{Constellation Optimization and Energy Difference Detection in Massive SIMO Systems With Rician Channels}}, 
  year={2025},
month={Aug.},
  volume={73},
  number={8},
  pages={6928-6942},
  doi={10.1109/TCOMM.2025.3540972}}

@ARTICLE{simo_nograss2,
  author={Mallik, Ranjan K.},
  journal={IEEE Trans. Wireless Commun.}, 
  title={{Multi-Level Signaling With Noncoherent SIMO Over a Rayleigh Channel}}, 
  year={2024},
month = {Nov.},
  volume={23},
  number={11},
  pages={16139-16154},
  doi={10.1109/TWC.2024.3437774}}

@ARTICLE{simo_grass2,
  author={Duong, Son T. and Nguyen, Ha H. and Bedeer, Ebrahim and Barton, Robert},
  journal={IEEE Trans. Wireless Commun.}, 
  title={{Design and Detection of Unitary Constellations in Non-Coherent SIMO Systems for Short Packet Communications}}, 
  year={2024},
  volume={23},
  number={10},
  pages={12873-12887},
  doi={10.1109/TWC.2024.3396721}}

@ARTICLE{expmap1,
  author={Kammoun, I. and Belfiore, J.-C.},
  journal={IEEE Commun. Lett.}, 
  title={{A New Family of Grassmann Space-Time Codes for Non-Coherent MIMO Systems}}, 
  year={2003},
  volume={7},
  number={11},
  pages={528-530},
  doi={10.1109/LCOMM.2003.820081}}

@ARTICLE{expmap2,
  author={Kammoun, Ines and Cipriano, Antonio M. and Belfiore, Jean-Claude},
  journal={IEEE Trans. Wireless Commun.}, 
  title={{Non-Coherent Codes over the Grassmannian}}, 
  year={2007},
month = {Oct.},
  volume={6},
  number={10},
  pages={3657-3667},
  doi={10.1109/TWC.2007.06059}}

@ARTICLE{scox,
  author={Schmidl, T.M. and Cox, D.C.},
  journal={IEEE Trans. Commun.}, 
  title={{Robust Frequency and Timing Synchronization for OFDM}}, 
  year={1997},
  month={Dec.},
  volume={45},
  number={12},
  pages={1613-1621},
  doi={10.1109/26.650240}}

@ARTICLE{mimo_tran,
  author={Hassibi, B. and Hochwald, B.M.},
  journal={IEEE Trans. Inf. Theory}, 
  title={{How Much Training is Needed in Multiple-Antenna Wireless Links?}}, 
  year={2003},
month={Apr.},
  volume={49},
  number={4},
  pages={951-963},
  doi={10.1109/TIT.2003.809594}}

@ARTICLE{gt2,
  author={Hochwald, B.M. and Marzetta, T.L.},
  journal={IEEE Trans. Inf. Theory}, 
  title={{Unitary Space-Time Modulation for Multiple-Antenna Communications in Rayleigh Flat Fading}}, 
  year={2000},
month={Mar.},
  volume={46},
  number={2},
  pages={543-564},
  doi={10.1109/18.825818}}

@ARTICLE{gt3,
  author={Lizhong Zheng and Tse, D.N.C.},
  journal={IEEE Trans. Inf. Theory}, 
  title={{Communication on the Grassmann Manifold: a Geometric Approach to the Noncoherent Multiple-Antenna Channel}}, 
  year={2002},
month={Feb.},
  volume={48},
  number={2},
  pages={359-383},
  doi={10.1109/18.978730}}

@ARTICLE{do1,
  author={Gohary, Ramy H. and Davidson, Timothy N.},
  journal={IEEE Trans. Inf. Theory}, 
  title={{Noncoherent MIMO Communication: Grassmannian Constellations and Efficient Detection}}, 
month = {Mar.},
  year={2009},
  volume={55},
  number={3},
  pages={1176-1205},
  doi={10.1109/TIT.2008.2011512}}

@ARTICLE{do2,
  author={Beko, Marko and Xavier, JoÃo and Barroso, Victor A. N.},
  journal={IEEE Trans. Signal Process}, 
  title={{Noncoherent Communication in Multiple-Antenna Systems: Receiver Design and Codebook Construction}}, 
  year={2007},
month={Dec.},
  volume={55},
  number={12},
  pages={5703-5715},
  doi={10.1109/TSP.2007.901151}}

@inproceedings{dp,
  author    = {J. Álvarez-Vizoso and D. Cuevas and C. Beltrán and I. Santamaria and V. Tuček and G. Peters},
  title     = {{Coherence-Based Subspace Packings for MIMO Noncoherent Communications}},
  booktitle = {2022 30th European Signal Processing Conference (EUSIPCO)},
  year      = {2022},
  pages     = {1906--1910},
  address   = {Belgrade, Serbia},
  publisher = {IEEE},
}

@inproceedings{struct_can,
  author={Cuevas, Diego and Beltrán, Carlos and Gutiérrez, Mikel and Santamaria, Ignacio and Tuček, Vít},
  booktitle={2024 IEEE 13rd Sensor Array and Multichannel Signal Processing Workshop (SAM)}, 
  title={{Structured Multi-Antenna Grassmannian Constellations for Noncoherent Communications}}, 
  year={2024},
  volume={},
  number={},
  pages={1-5},
address = {Corvallis, OR, USA},
  doi={10.1109/SAM60225.2024.10636457}}

@ARTICLE{ub,
  author={Cuevas, Diego and Álvarez-Vizoso, Javier and Beltrán, Carlos and Santamaria, Ignacio and Tuček, Vít and Peters, Gunnar},
  journal={IEEE Trans. Commun.}, 
  title={{Union Bound Minimization Approach for Designing Grassmannian Constellations}}, 
  year={2023},
month={Apr.},
  volume={71},
  number={4},
  pages={1940-1952},
  doi={10.1109/TCOMM.2023.3244965}}

@ARTICLE{grassma,
  author={Álvarez-Vizoso, Javier and Cuevas, Diego and Beltrán, Carlos and Santamaria, Ignacio and Tuček, Vít and Peters, Gunnar},
  journal={IEEE Trans. Inf. Theory}, 
  title={{Constrained Riemannian Noncoherent Constellations for the MIMO Multiple Access Channel}}, 
  year={2023},
month={July},
  volume={69},
  number={7},
  pages={4559-4578},
  doi={10.1109/TIT.2023.3249903}}

\end{document}